\documentclass[journal,twoside]{IEEEtran}
\usepackage{amsmath,amsfonts}
\usepackage{algorithmic}
\usepackage{algorithm}
\usepackage{array}
\usepackage[caption=false,font=normalsize,labelfont={footnotesize},textfont={footnotesize}]{subfig}
\usepackage{textcomp}
\usepackage{stfloats}
\usepackage{url}
\usepackage{verbatim}
\usepackage{graphicx}
\usepackage{cite}
\usepackage[hidelinks]{hyperref}
\usepackage{orcidlink}
\usepackage{multirow}
\usepackage{mathtools}

\usepackage{color}
\usepackage{import}

\newcommand{\shrt}[1]{\hspace*{1.5pt}{#1}\hspace*{1.5pt}}
\newcommand{\?}{\hspace*{-0.5pt}}

\newcommand{\m}[1]{\mathbf{#1}} 
\newcommand{\mi}[1]{\mathbf{#1}^{-1}} 
\newcommand{\mh}[1]{\mathbf{#1}^{\mathrm{H}}} 

\newcommand{\lt}[1]{_{\text{#1}}}

\newcommand{\norm}[1]{\| #1 \|}

\newcommand{\rmH}{^\mathrm{H}}

\DeclareMathOperator{\Ex}{\mathbb{E}}

\newcommand{\yy}[0]{\lt{yy}}
\newcommand{\xx}[0]{\lt{xx}}
\newcommand{\vv}[0]{\lt{vv}}
\newcommand{\p}[0]{\m{p}}
\newcommand{\ps}[0]{\m{p}\lt{s}}

\newcommand{\tr}[0]{\operatorname{tr}}
\newcommand{\numero}{{\fontfamily{qpl}\textnumero}\,}

\newcommand{\srp}[0]{P}

\newcommand{\fig}[0]{Fig.\;}
\newcommand{\tab}[0]{Tab.\;}
\newcommand{\sect}[0]{Section~}
\newcommand{\eq}[0]{}

\newcommand{\mrk}[1]{\textit{#1}}

\begin{document}

\title{A Steered Response Power Method for Sound Source Localization with Generic Acoustic Models}

\author{%
	Kaspar~M\"uller,
	\thanks{%
		K.~M\"uller, M.~Buck, and T.~Wolff are with the Audio AI R\&D Department, Cerence AI, 89077 Ulm, Germany (e-mail:
		\{\href{mailto:kaspar.mueller@cerence.com}{kaspar.mueller},
		\href{mailto:markus.buck@cerence.com}{markus.buck},
		\href{mailto:tobias.wolff@cerence.com}{tobias.wolff}\}@cerence.com).}%
	Markus~Buck,%
	~\IEEEmembership{Member,~IEEE,}
	Simon~Doclo,
	~\IEEEmembership{Senior~Member,~IEEE,}
	\thanks{%
		S.~Doclo is with the Department of Medical Physics and Acoustics, and the Cluster of Excellence Hearing4all, University of Oldenburg, 26129 Oldenburg, Germany (e-mail: \href{mailto:simon.doclo@uni-oldenburg.de}{simon.doclo@uni-oldenburg.de}).}%
	Jan~\O{}stergaard,
	~\IEEEmembership{Senior~Member,~IEEE,}
	\thanks{%
		J.~{\O}stergaard is with the Department of Electronic Systems, Aalborg University, 9220 Aalborg East, Denmark (e-mail: 
		\href{mailto:jo@es.aau.dk}{jo@es.aau.dk}).}%
	and Tobias~Wolff
	\thanks{%
		This project has received funding from the SOUNDS European Training Network -- an European Union’s Horizon 2020 research and innovation programme under the Marie Skłodowska-Curie grant agreement No. 956369.}%
	\vspace*{-0.5\baselineskip}%
}

\markboth{}
{M\"uller \MakeLowercase{\textit{et al.}}: A SRP Method for Sound Source Localization with Generic Acoustic Models}

\IEEEpubid{%
	\begin{minipage}{0.82\textwidth}\ \\[24pt]
		© 2025 IEEE. Personal use of this material is permitted. Permission from IEEE must be obtained for all other uses, in any current or future media, including reprinting/republishing this material for advertising or promotional purposes, creating new collective works, for resale or redistribution to servers or lists, or reuse of any copyrighted component of this work in other works.
	\end{minipage}
}

\maketitle

\begin{abstract}

The steered response power (SRP) method is one of the most popular approaches for acoustic source localization with microphone arrays.
It is often based on simplifying acoustic assumptions, such as an omnidirectional sound source in the far field of the microphone array(s), free field propagation, and spatially uncorrelated noise.
In reality, however, there are many acoustic scenarios where such assumptions are violated.
This paper proposes a generalization of the conventional SRP method that allows to apply generic acoustic models for localization with arbitrary microphone constellations.
These models may consider, for instance, level differences in distributed microphones, the directivity of sources and receivers, or acoustic shadowing effects.
Moreover, also measured acoustic transfer functions may be applied as acoustic model.
We show that the delay-and-sum beamforming of the conventional SRP is not optimal for localization with generic acoustic models.
To this end, we propose a generalized SRP beamforming criterion that considers generic acoustic models and spatially correlated noise, and derive an optimal SRP beamformer.
Furthermore, we propose and analyze appropriate frequency weightings.
Unlike the conventional SRP, the proposed method can jointly exploit observed level and time differences between the microphone signals to infer the source location.
Realistic simulations of three different microphone setups with speech under various noise conditions indicate that the proposed method can significantly reduce the mean localization error compared to the conventional SRP and, in particular, a reduction of more than 60\% can be archived in noisy conditions.
\end{abstract}


\section{Introduction}

\begin{figure}[!t]
	\centering%
	\vspace*{-2mm}%
	\hspace*{1mm}%
	\subfloat[%
	Conventional SRP application: 
	one or multiple distributed microphone arrays are in the far field of a sound source.%
	]{%
		\includegraphics[width=0.45\linewidth, trim=2mm 1mm 1mm 2mm, clip]{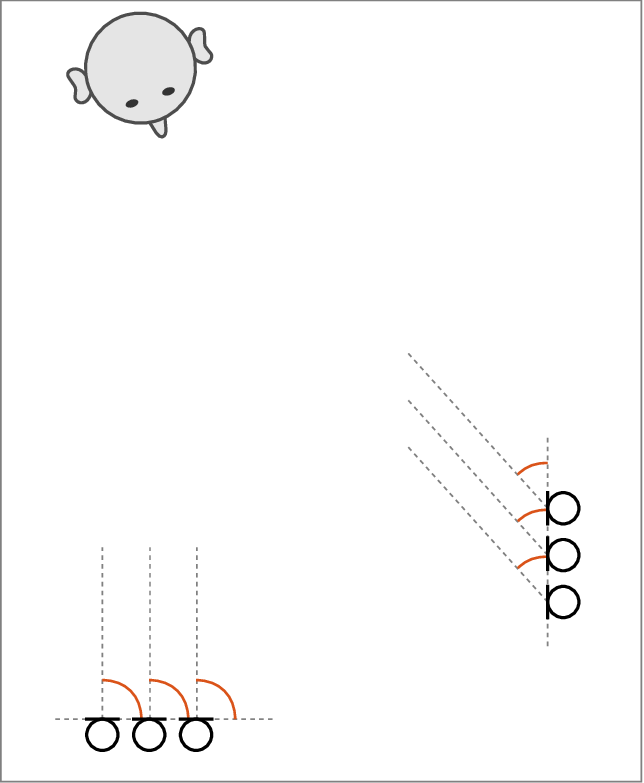}%
		\label{fig:SRP_setup_farfield}}
	\hfill
	\vrule height 48.5mm
	\hfill
	\subfloat[%
	Exemplary generic application: distributed microphones or arrays are in the near field of a source and might be affected by shadowing effects and different directivities of the source or microphones.%
	]{%
		\includegraphics[width=0.45\linewidth, trim=2mm 1mm 1mm 2mm, clip]{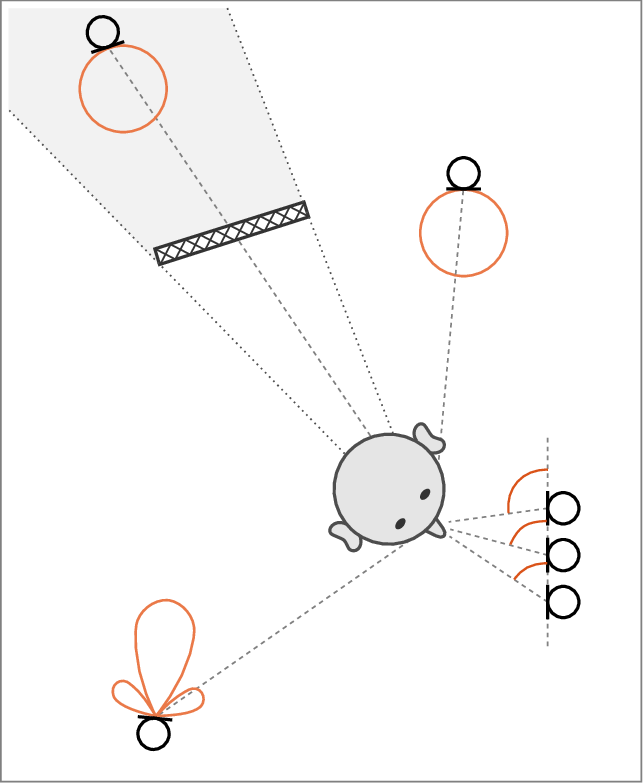}%
		\label{fig:SRP_setup_nearfield}}
	\hspace*{1mm}%
	\caption{%
		Exemplary microphone setups for source localization. Unlike in~(a), the microphone level differences in (b) contain relevant source position cues.%
	}
	\label{fig:SRP_setup_comparison}
\end{figure}

\IEEEPARstart{A}{coustic} sound source localization is a frequently required task.
Beyond source location estimation itself, it is fundamental for various applications, such as teleconferencing, adaptive beamforming, speaker separation, autonomous driving, or robotics.
The steered response power (SRP) method is one of the most commonly used methods for source localization with microphone arrays.
Its conceptual idea is to use a steered beamformer in order to scan the space for a sound source by observing the steering direction (or position) with maximum beamformer output power.
In the 1990s, Omologo and Svaizer observed that using the phase information of microphone cross power spectra is a useful strategy for time-difference-of-arrival-based source localization~\cite{Omologo1994, Omologo1997}.
This idea of exploiting microphone cross power spectral densities by means of a steered delay-and-sum beamformer further evolved to the SRP method in its current form as a source localization standard \cite{DiBiase2001, Madhu2008}.
In particular, the variant applying the phase transform (PHAT)~\cite{Knapp1976} in order to use the cross power spectrum phase asserted itself as the popular SRP-PHAT that is known for its robustness against reverberation~\cite{Brandstein1997a, DiBiase2001, Chen2006}.
We refer to the standard SRP (including SRP-PHAT) as conventional SRP (CSRP).
An extensive literature study reviewing its background and presenting various extensions has been published recently~\cite{Grinstein2024}.
The CSRP is based on ideal, simplifying acoustic assumptions, namely an omnidirectional point source in the far field of the microphones, free-field propagation and spatially uncorrelated noise~\cite{Madhu2008}. In practice, these assumptions are usually not met~\cite{Gannot2019}.
Nevertheless, SRP-PHAT usually performs well with microphone arrays in the far field of a source.
Hence, the CSRP is typically used with microphone arrays whose aperture is much smaller than the distance to the sound source -- either with a single array for direction-of-arrival (DOA) estimation, or with multiple distributed microphone arrays for source position estimation.
In the latter case, the CSRP is often processed for each array individually and weighted and summed up afterwards, or triangulation is used to infer the source location~\cite{Aarabi2003, Brutti2006, Plinge2014}.
In various other setups, however, the simple acoustic assumptions are violated to a greater extent.
Typical examples are setups with distributed microphones in the near field of a source including the emerging field of wireless acoustic sensor networks, or setups involving directional microphones or sources.
Another common application is binaural source localization, for instance with hearing aid devices, where the acoustic head shadow causes frequency-dependent interaural level and time differences that significantly deviate from free-field propagation~\cite{Rohdenburg2008, Raspaud2010}.
In such scenarios, not only phase differences between microphone power spectra can be used to determine the source location but also the power differences carry relevant source location cues.
However, power differences are not exploited by the CSRP (see  \fig\ref{fig:SRP_setup_comparison}).
\IEEEpubidadjcol

Various approaches have been proposed to improve the localization performance of SRP-PHAT, for instance, 
with setups involving distributed microphones or sensor networks \cite{Hummes2011, Marti2011, Huang2021, Cakmak2022, Cakmak2024}, by alternative prewhitening \cite{He2018, Wang2024}, or by exploiting an auxiliary microphone \cite{Bruemann2024}.
Other recent contributions focus on the problem of SRP localization of multiple sound sources~\cite{Dang2024, Lai2024, Tengan2024}.
Furthermore, several extension or alternative grid search methods have been proposed to reduce the computational complexity~\cite{Zotkin2004, Dmochowski2007, Do2007, Cobos2011a, Dietzen2021}.
There are also approaches using measured head-related transfer functions or head models for binaural steered-beamforming localization \cite{Rohdenburg2008, Zohourian2018}.
Alternative SRP approaches or extensions involving machine learning are proposed, for instance, in
\cite{Salvati2016, Salvati2016a, Pertila2017, Grinstein2023, Grinstein2023a}
to improve localization performance under realistic acoustic conditions.
However, to the best of our knowledge there is no signal-processing-based work addressing generic acoustic conditions in general by involving more complex acoustic propagation models and noise characteristics for SRP-based localization with arbitrary microphone constellations.

In this paper, we propose a generalization of the conventional frequency-domain SRP method with regard to the just mentioned aspects:
The presented method allows to apply advanced, setup-specific acoustic propagation models.
We show that this enables a joint exploitation of phase and level information for localization.
Moreover, arbitrary noise characteristics can be taken into account to improve robustness against spatially correlated or inhomogeneous noise.
In particular, it is shown that simply replacing the free far-field model of the CSRP by other acoustic models is not optimal with standard SRP beamformers.
In order to overcome this limitation, we propose a generalized steered response power (GSRP) beamformer design under consideration of generic acoustic models and noise fields and derive corresponding GSRP beamformers.
Moreover, we propose and analyze frequency weightings of the GSRP beamformer output as potential alternatives to the PHAT weighting.
The presented methods are evaluated in realistic simulations of different scenarios.

The paper is structured as follows:
In \sect\ref{sec:SRP}, we review the generic mathematical SRP framework and the conventional SRP with its typical applications. 
\sect\ref{sec:SRP_extension} introduces the idea behind the generalization of the conventional SRP and shows limitations of standard SRP beamformers in combination with generic acoustic models.
The GSRP beamformer design, the resulting beamformers, and frequency weightings are presented in \sect\ref{sec:GSRP}.
The proposed methods are evaluated in comparison to the conventional SRP in \sect\ref{sec:evaluations}.

\section{The Steered Response Power Method}
\label{sec:SRP}

\noindent
This section introduces the signal model and reviews the generic mathematical framework of the SRP method for acoustic source localization\footnote{%
	Note that in this work, source localization explicitly refers to both the DOA estimation and the source position estimation.
	Throughout this work, we use the more general source position $\ps$ as the quantity to estimate from which the source DOA $\theta\lt{s}$ can be derived.}
as well as its most common variant which we refer to as conventional SRP (CSRP).

\subsection{Signal Model in the Frequency Domain}
\label{sec:signal_model}

\noindent
Throughout this work, we assume that there is only one active target sound source at position $\ps \shrt{\in} \mathbb{R}^3$.
The noisy sound is recorded by $M$ microphones at different microphone positions~$\p_m$, $m \shrt{\in} \{1, \dots, M\}$.
The microphone signals in the frequency domain at angular frequency $\omega \shrt{=} 2 \pi\! f$ are given by
\begin{align}
	\m{y}(\omega)
	&= 
	\m{x}(\omega) + \m{v}(\omega)
	=
	\m{h}\lt{s}(\omega) \, s(\omega) + \m{v}(\omega) \,,
\end{align}
where the  vector $\m{y}(\omega) \shrt{=} [y_1(\omega), ..., y_M(\omega)]^\mathrm{T}$ comprises the $M$ complex-valued microphone signals which consist of the noise-free microphone signals~$\m{x}(\omega)$ and the noises~$\m{v}(\omega)$.
Furthermore,
$\m{h}\lt{s}(\omega) \shrt{=} \m{h}(\omega, \ps) \shrt{=} [h_1(\omega, \ps), ..., h_M(\omega, \ps)]^\mathrm{T}$ is the vector of acoustic transfer functions (ATFs) from the source position~$\ps$ to the microphone positions~$\p_1, ..., \p_M$, and $s(\omega)$ is the clean source signal.
The ATFs~$\m{h}\lt{s}(\omega)$ explicitly include all effects of sound propagation, such as reverberation, the distance-dependent attenuation, shadowing effects and the directivity of the source or the microphones.
Additionally, we assume that the source signal and noise signals are independent with zero mean and thus
the signal covariance matrix\footnote{Often also called {spatial covariance matrix} or {spatial correlation matrix}.} (SCM)
\begin{align}
	\m\Phi\yy(\omega)
	&= 
	\Ex \{ \m{y}(\omega) \, \mh{y}(\omega) \}
	= 
	\m\Phi\xx(\omega) + \m\Phi\vv(\omega) \,,
	\label{eq:Phi_yy=Phi_xx+Phi_vv}
\end{align}
can be expressed as the sum of the noise-free SCM
\begin{align}
	\m\Phi\xx(\omega)
	&=
	\Ex \{ \m{x}(\omega)  \, \mh{x}(\omega) \}
	=
	\varPhi_{ss}(\omega) \, \m{h}\lt{s}(\omega) \, \mh{h}\lt{s}(\omega)
	\label{eq:Phi_xx_rank1}
\end{align}
and noise\;covariance\;matrix\,(NCM) $\m\Phi\vv(\omega) {=} \Ex \{ \m{v}(\omega) \mh{v}(\omega) \!\}$,
where $\Ex\{\cdot\}$ is the expectation operator,
$\{\cdot\}\rmH$ denotes the Hermitian conjugate, and
$\varPhi_{ss}(\omega) \shrt{=} \Ex \{ |\m{s}(\omega)|^2 \}$ is the power spectral density (PSD) of the source signal.
The noise-free SCM in~\eq\eqref{eq:Phi_xx_rank1} is a rank-one matrix when there is a single sound source.
Besides, we assume that $\m\Phi\vv(\omega)$ is always invertible.

\subsection{Generic SRP Framework and Conventional SRP}
\label{sec:SRP_framework}

\begin{figure}[tp]
	\centering
	\vspace*{3mm}%
	\def\svgwidth{0.98\linewidth}
	\hspace*{-1mm}%
	\footnotesize
	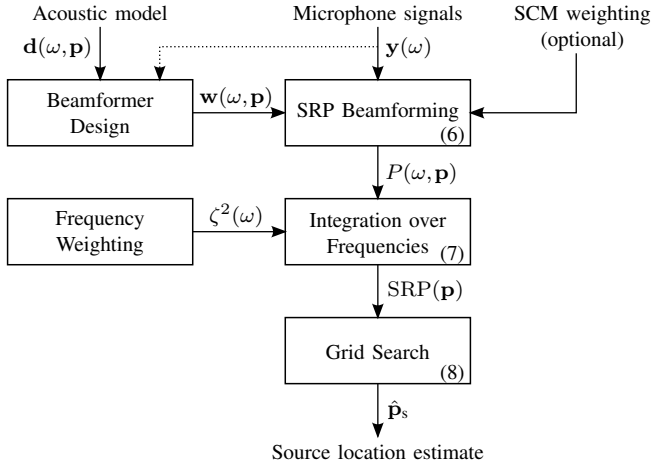%
	\caption{Structure of the generic steered response power framework.}%
	\label{fig:SRP_structure}%
\end{figure}

\noindent
The structure of the generic mathematical SRP framework is shown in \fig\ref{fig:SRP_structure}.
In SRP, a beamformer is steered towards multiple candidate points to scan the space for a sound source.
The beamformer output for point~$\p$ is
\begin{align}
	z(\omega,\p)
	&=
	\mh{w}(\omega, \p) \; \m{y}(\omega)
\end{align}
with beamformer weights $\m{w}(\omega,\p) \shrt{=} [w_1(\omega,\p),\, ...,\, w_M(\omega,\p)]^\mathrm{T}$.
The typical beamformer in the CSRP is a delay-and-sum (DS) beamformer~\cite{DiBiase2001, Madhu2008} with\footnote{In some publications, e.g., in \cite{VanTrees2002}, the DS beamformer weights are scaled by $1/M$. This, however, does not affect the SRP localization result.}
\begin{align}
	\m{w}\lt{DS}(\omega, \p) 
	&=
	\m{d}\lt{ff}(\omega, \p)
	=
	\big[
	e^{-j \omega T_1(\p)}, \dots, e^{-j \omega T_M(\p)} 
	\big]^\mathrm{T} \,,
	\label{eq:free_far_field}
\end{align}
where 
$\m{d}\lt{ff}(\omega, \p)$ is the acoustic free far-field model that models the time of flight
$T_m(\p)$ of the actual, unknown ATF $\m{h}(\omega, \p)$ from a position~$\p$ to each microphone position~$\p_m$.
However, there are also publications proposing other beamformer approaches for SRP, such as the minimum power distortionless response\footnote{%
	Note that the authors of \cite{Dmochowski2010, Salvati2016} refer to it as minimum variance distortionless response (MVDR) beamformer.
	However, we use the denotation MPDR to distinguish from MVDR as also done, for instance, in \cite{VanTrees2002}.} (MPDR) beamformer~\cite{Dmochowski2010, Salvati2016}%
, target beamforming or null-steering beamforming~\cite{Zohourian2018}.
The PSD of the SRP beamformer output (i.e., the \textit{steered response power}) is
\begin{align}
	\hspace*{-6pt}\srp(\omega, \p)
	&=
	\Ex \{ |z(\omega, \p)|^2 \}
	=
	\mh{w}(\omega, \p) \, \m\Phi\yy(\omega) \, \m{w}(\omega, \p) \,.
	\label{eq:Phi_zz}%
\end{align}
In addition, various contributions, e.g.,~\cite{Bedard1994, Brandstein1997a, DiBiase2001, Valin2007, Abutalebi2011, Braun2015}, propose a weighting of the SCM $\m\Phi\yy(\omega)$.
The most common weighting in the CSRP is the phase transform (PHAT) weighting~\cite{Knapp1976}
which scales each element of the SCM to magnitude one.
In combination with the DS beamforming, this weighting yields the popular SRP-PHAT which is known for its robustness against reverberation~\cite{Brandstein1997a, DiBiase2001, Chen2006, Zhang2008}.
A broadband SRP value is determined by integrating the SRP beamformer PSDs $P(\omega, \p)$ over all frequencies:
\begin{align}
	\operatorname{SRP}(\m{p})
	&=
	\frac{1}{2 \pi}
	\int\limits_{-\infty}^{\infty}
	\zeta^2(\omega) \;
	\srp(\omega, \p)  \; d \omega \,,
	\label{eq:SRP(p)}
\end{align}
where $\zeta^2(\omega)$ is a generic, real-valued, positive frequency weighting factor.
Finally, the SRP is computed at a grid of possible source locations (which we call SRP map) and the source position estimate is the position with maximum SRP output~\eq\eqref{eq:SRP(p)}, i.e.,
\begin{align}
	\hat{\p}\lt{s}
	&=
	\arg \max_{\p} \; \operatorname{SRP}(\p) \,.
	\label{eq:grid_search}
\end{align}%

\section{SRP with Generic Acoustic Models}
\label{sec:SRP_extension}

\noindent
The CSRP is not capable of exploiting microphone signal level differences for localization and extracts all source position cues purely from the observed TDOAs between the microphones.
This is mainly because the CSRP is based on the acoustic free far-field model of~\eq\eqref{eq:free_far_field} which purely models delays.
But also the PHAT weighting removes level information of the microphone covariances so that only phase differences remain to determine the source location.
While this is suitable in typical CSRP applications with microphone arrays in the far field of the sound source, 
exploiting level differences might be highly desirable in generic microphone setups (cf. \fig\ref{fig:SRP_setup_comparison}).
This can be achieved by considering both level and phase information in the acoustic model.
For instance, for sound source localization with distributed microphones, the acoustic free near-field model~\cite{Kuttruff2016} can be beneficial.
It is given by
\begin{align}
	\m{d}_{\text{nf}}(\omega, \p)
	&=
	\bigg[
	\frac{1}{4 \pi r_1\!(\p)} \, e^{-j \omega T_1\!(\p)}, \dots, \frac{1}{4 \pi r_{\!M}\!(\p)} \, e^{-j \omega T_{\!M}\!(\p)}
	\bigg]^\mathrm{T} \hspace*{-3pt}
	\label{eq:free_near_field}
\end{align}
with $T_m(\p) \shrt{=} r_m(\p) / c$,
where $r_m(\p) \shrt{=} \norm{\p - \p_m}$ is the distance between $\p$ and microphone $m$,
and $c$ is the speed of sound.
It exploits that the signals in microphones in the vicinity of the sound source have a significantly higher level than in distant microphones.
Moreover, one can reduce the mismatch between the acoustic model and the actual acoustical conditions by using generic acoustic models for SRP.
For example, for binaural source localization with hearing aid microphones, using measured or modeled head-related transfer functions (HRTFs) as acoustic model is preferable over a free-field model.
This is because by using HRTFs the acoustic shadowing effect of the head is taken into account which causes 
frequency-dependent interaural level and time differences that significantly differ from free-field assumptions, especially for lateral sound sources~\cite{Rohdenburg2008, Raspaud2010}.
However, considering generic acoustic models for SRP is not straightforward, as is shown with the following example.

\begin{figure}[!t]
	\centering
	\vspace*{-3mm}%
	\subfloat[%
	SRP with DS beamforming.
	]{%
		\includegraphics[scale=0.35, trim=5mm 0mm 20mm 8mm, clip]{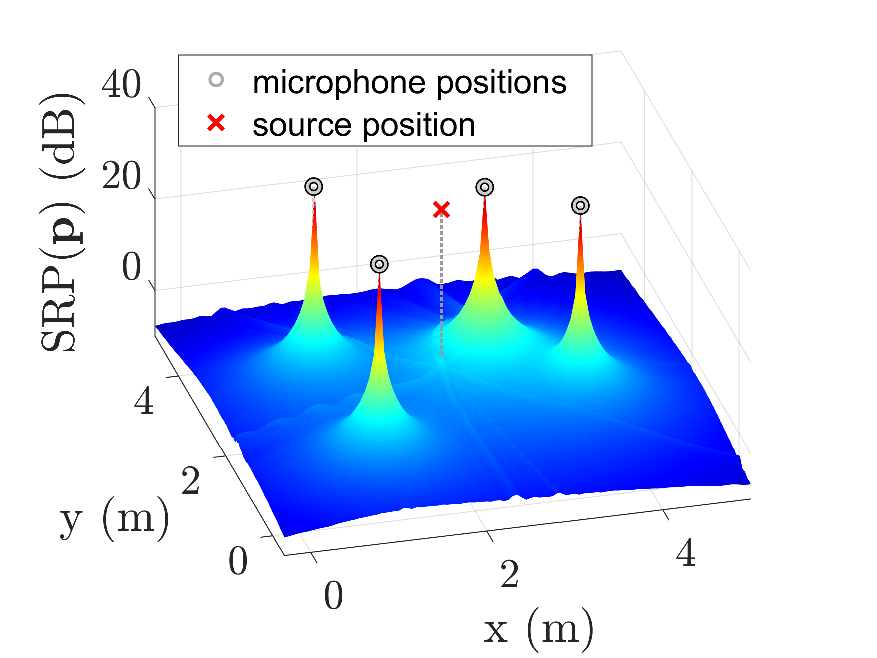}%
		\label{fig:SRP_DS_nf}}%
	\hfill
	\subfloat[%
	SRP with MVDR beamforming.
	]{%
		\includegraphics[scale=0.35, trim=1mm 0mm 20mm 8mm, clip]{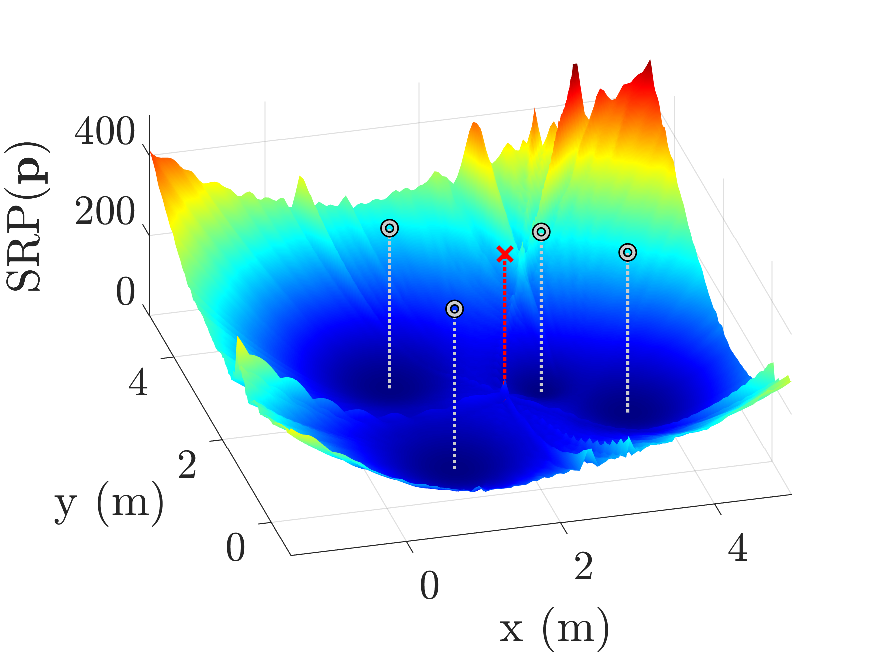}%
		\label{fig:SRP_MVDR_nf}}%
	\caption{
		SRP maps when employing the acoustic free near-field model of~\eq\eqref{eq:free_near_field} with a DS~\eqref{eq:free_far_field} or MVDR~\eqref{eq:w_MVDR_I} beamformer.
		The simulated sound source (red cross) is located between four distributed microphones (gray circles).
		In this exemplary scenario, SRP localization with a DS or MVDR beamformer fails.
	}
	\label{fig:DS_MVDR_nf}
\end{figure}

Let us consider a simple simulated scenario with four distributed microphones in the free field (no reverberation).
An omnidirectional sound source in the middle of the four microphones emits white noise (desired signal).
In order to consider near-field effects, it would be intuitive to simply apply the free near-field model of~\eq\eqref{eq:free_near_field} instead of the free far-field model of~\eq\eqref{eq:free_far_field}
in the conventional DS beamforming formulation of~\eq\eqref{eq:free_far_field}, i.e., $\m{w}\lt{DS,nf}(\omega, \p) \shrt{=} \m{d}\lt{nf}(\omega, \p)$.
However, we can see that the magnitude of $\m{d}_{\text{nf}}(\omega, \p)$ in~\eq\eqref{eq:free_near_field} goes to infinity if $\p$ approaches one of the microphone positions~$\p_m$, since $r_m(\p)$ in the denominator of the $m$-th element of the vector in~\eq\eqref{eq:free_near_field} goes to zero and thus
\begin{align}
	\lim_{\p \rightarrow \p_m} \!\!
	\Vert \m{d}_{\text{nf}}(\omega, \p) \Vert
	&=
	\infty\;\;\;
	\forall \; m \in \{1, \dots, M\} \,.
	\label{eq:d_nf_infty}%
\end{align}
Conversely, we can see that the magnitude of $\m{d}_{\text{nf}}(\omega, \p)$ diminishes for an increasing distance~$r_m(\p) \shrt{\rightarrow} \infty$ between $\p$ and all microphone positions~$\p_m$:
\begin{align}
	\lim_{r_m(\p) \rightarrow \infty} \!\!
	\Vert \m{d}_{\text{nf}}(\omega, \p) \Vert 
	&=
	0 \,\;\;\;
	\forall \; m \in \{1, \dots, M\} \,.
	\label{eq:d_nf_0}%
\end{align}%
With~\eq\eqref{eq:Phi_zz}, we see that the near-field DS beamformer output power
$\srp\lt{DS,nf}(\omega,\p) \shrt{=} \mh{d}\lt{nf}(\omega, \p) \, \m\Phi\yy(\omega)\, \m{d}\lt{nf}(\omega, \p)$ 
also goes to infinity if $\p$ gets close to a microphone position~$\p_m$, and to zero for very distant positions, respectively.
This behavior is visualized in \fig\ref{fig:SRP_DS_nf} (in logarithmic scale).

Hence, it becomes obvious that the DS beamformer would require an appropriate normalization.
Let us therefore consider an SRP beamformer with a distortionless response constraint, such as MVDR or MPDR, which normalizes the beamformer output PSD so that it equals the source PSD~\cite{VanTrees2002}.
Under the assumption of spatially uncorrelated and homogeneous noise, the NCM and its inverse are diagonal matrices with
\begin{align}
	\m\Phi\vv(\omega)
	&=
	{\sigma_v^2}(\omega) \; \m{I} \,,\;
	\text{ and } \;
	\mi\Phi\vv(\omega)
	=
	\frac{1}{{\sigma_v^2}(\omega)} \; \m{I} \,,
	\label{eq:Phi_vv_IID}
\end{align}
where $\sigma_{v}^2(\omega)$ is the noise power and $\m{I}$ is the identity matrix.
With this, the MVDR beamformer simplifies to
\begin{align}
	\m{w}\lt{MVDR}(\omega, \p)
	=
	\frac{\mi\Phi\vv(\omega)  \, \m{d}(\omega, \p)}{
		\mh{d}(\omega, \p) \, \mi\Phi\vv(\omega) \, \m{d}(\omega, \p)}
	&\stackrel{\eqref{eq:Phi_vv_IID}}{=}
	\frac{\m{d}(\omega, \p)}{ {\norm{\m{d}(\omega, \p)}}^2 } \,,
	\label{eq:w_MVDR_I}
\end{align}
which can be interpreted as a normalized DS beamformer.
With~\eq\eqref{eq:d_nf_infty} and~\eqref{eq:d_nf_0}, we can see that the MVDR beamformer weights of~\eq\eqref{eq:w_MVDR_I} behave in the opposite way to the previous DS example: The magnitude of the MVDR weights (and thus also its output PSD) goes to zero if $\p$ approaches a microphone position~$\p_m$, while it goes to infinity for increasingly distant positions.
It can be shown that this also holds for MPDR and for MVDR with arbitrary (invertible) NCMs.
This behavior is visualized in \fig\ref{fig:SRP_MVDR_nf}.

This simple example shows that a correct SRP localization with DS or a distortionless response beamformer, such as MVDR, is not guaranteed with a near-field model -- even in the noise-free case with no mismatch between the acoustic model and the simulated ATFs, i.e., $\m{d}(\omega,\p) \shrt{=} \m{h(\omega,\p)}$. 
In fact, this is not an exclusive problem of the near-field model but rather it affects any acoustic model that incorporates source-position-dependent level information and not purely phase information.
Moreover, also with a far-field model that ignores level information, the influence of noise on the localization result cannot be controlled with the conventional SRP.
For these reasons, we propose a novel beamforming method for SRP in the following section that addresses the above mentioned problems.

\section{Generalized Steered Response Power Method}
\label{sec:GSRP}

\noindent
In this section, a generalized steered response power (GSRP) method for sound source localization is introduced.
It generalizes the conventional SRP (see \sect\ref{sec:SRP_framework}) in the sense that it allows for using generic acoustic models by means of a specific beamformer design to avoid the effects shown in the previous section (see \fig\ref{fig:DS_MVDR_nf}). 
Furthermore, generic inhomogeneous noise characteristics are taken into account.
In addition, we propose appropriate frequency weightings and briefly address some practical aspects of GSRP.

\subsection{GSRP Beamformer Design Criteria}
\label{sec:GSRP_bf_design}

\noindent
In order to deduce an appropriate GSRP beamformer design, we start from the following obvious SRP localization objective:
The SRP map $\operatorname{SRP}(\p)$ must have a unique global maximum at the source position $\ps$ so that $\arg \max_{\p} \operatorname{SRP}(\p) \shrt{=} \ps$.
With~\eq\eqref{eq:SRP(p)}, this objective can be translated into a narrowband equivalent:
The SRP beamformer output PSD~$\srp(\omega, \p)$ shall have a global maximum at the source position $\p \shrt{=} \ps$ at each frequency~$\omega$, i.e.,
\begin{align}
	\srp(\omega,\ps) \geq \srp(\omega,\p) \;\;\;
	\forall \; \ps, \; \p, \; \omega\,.
	\label{eq:objective_Phi_zz}
\end{align}
Note that we accept that the global maximum might not be unique in each frequency $\omega$ which can be, for instance, due to spatial aliasing.
However, if multiple global maxima of $\srp(\omega,\p)$ exist, their respective locations are frequency-dependent and thus vary over frequencies except for the commonly required,  frequency-independent maximum at $\ps$.
Therefore, integrating over all frequencies in~\eq\eqref{eq:SRP(p)} usually leads to a unique global SRP maximum at $\ps$.

Under the assumption of statistically independent source and noise signals (cf.~\eq\eqref{eq:Phi_yy=Phi_xx+Phi_vv}),
we can split the beamformer output PSD of~\eq\eqref{eq:Phi_zz} into two PSD terms~$\srp^{(\m{x})}(\omega, \p)$ and~$\srp^{(\m{v})}(\omega, \p)$, which we refer to as \mrk{source response} and \mrk{noise response}, respectively:
\begin{flalign}
	\srp&(\omega, \p) 
	\label{eq:Phi_zz_sum}
	&
	\\
	&= 
	\underbrace{\mh{w}(\omega, \p) \, \m\Phi\xx(\omega) \, \m{w}(\omega, \p)}_{
		\text{\textit{source response} }
		\srp^{(\m{x})}(\omega, \p)} \,+\,
	\underbrace{\mh{w}(\omega, \p) \, \m\Phi\vv(\omega) \, \m{w}(\omega, \p)}_{
		\text{\textit{noise response} }
		\srp^{(\m{v})}(\omega, \p)} \,.
	\nonumber%
\end{flalign}
We treat these two terms separately in the following GSRP beamformer design.
Let us first consider the noise-free case, i.e., we only consider the source response $\srp^{(\m{x})}(\omega, \p)$.
The condition of~\eq\eqref{eq:objective_Phi_zz} then requires that the source response $\srp^{(\m{x})}(\omega, \p)$ has a global maximum at $\p \shrt{=} \ps$ at each frequency.
This defines a first GSRP beamformer design criterion as
\begin{equation}
	\boxed{%
		\;
		\begin{gathered}
			\text{\textbf{Criterion \numero1}}
			\\[-3pt]
			\text{\textit{max.~source response}}
		\end{gathered}
		\quad\;
		\begin{gathered}
			\srp^{(\m{x})}(\omega, \ps)
			\geq
			\srp^{(\m{x})}(\omega, \p)
			\\
			\forall \, \p, \, \ps, \;
			\text{with }
			\varPhi_{ss}(\omega) \shrt{>} 0\,.
		\end{gathered}
		\,
		\label{eq:criterion_no1}
	}
\end{equation}
We refer to this as the \mrk{maximum source response criterion}, where $\varPhi_{ss}(\omega) \shrt{>} 0$ is to avoid the trivial solution of
$\srp^{(\m{x})}(\omega, \ps) \shrt{=} \srp^{(\m{x})}(\omega, \p) \shrt{=} 0$.

Now we take into account the noise response~$\srp^{(\m{v})}(\omega, \p)$ in~\eq\eqref{eq:Phi_zz_sum}.
We propose a second beamformer design criterion that requires that the noise response is constant for all $\p$:
\begin{equation}
	\boxed{%
		\;
		\begin{gathered}
			\text{\;\textbf{Criterion \numero2}}
			\\[-3pt]
			\text{\textit{constant noise response}}
		\end{gathered}
		\quad\;
		\begin{gathered}
			\srp^{(\m{v})}(\omega, \p) = C \;\;\,
			\forall \; \p \,,
			\\
			\forall \, \text{ invertible } \m\Phi\vv(\omega)\,,
		\end{gathered}
		\;
		\label{eq:criterion_no2}
	}
\end{equation}
where $C$ is a positive, real-valued constant.
This \mrk{constant noise response} criterion implies that the SRP map in the noise-only case shall be flat.
It is intuitive that the condition of~\eq\eqref{eq:objective_Phi_zz} is fulfilled if both GSRP beamformer design criteria \eq\eqref{eq:criterion_no1} and \eq\eqref{eq:criterion_no2} are met because the maximum source response criterion ensures a global maximum of $\srp^{(\m{x})}(\omega, \p)$ at position~$\ps$ and the constant noise response criterion yields a position-indepen-dent, constant offset of the beamformer output PSD in~\eq\eqref{eq:Phi_zz_sum}.

\subsection{GSRP Beamformer Derivation: The MVCNR Beamformer}
\label{sec:GSRP_derivation_MVCNR}

\noindent
We consider the following generic linear beamformer formulation as approach for the GSRP beamformer derivation:
\begin{align}
	\m{w}(\omega, \p)
	&=
	\alpha(\omega, \p) \, \m{A} \, \m{h}(\omega, \p) \,,
	\label{eq:w_general_fomulation}
\end{align}
where $\m{A}$ is a complex-valued Hermitian, positive-definite matrix and $\alpha(\omega, \p)$ is a complex-valued scaling factor.
Specifically,~\eq\eqref{eq:w_general_fomulation} is a generalized formulation of optimum beamformers including maximum signal-to-noise ratio (max-SNR), minimum mean-square error (MMSE), MVDR and MPDR beamformers~\cite{VanVeen1988, VanTrees2002}, but it also includes general data-independent beamformers~\cite{VanVeen1988} such as DS beamformers.
We choose this approach from the family of optimum beamformers because the beamformers that are commonly used in the SRP context (cf.\ \sect\ref{sec:SRP_framework}) can also be assigned to this family.
Hence, the beamformer of~\eq\eqref{eq:w_general_fomulation} can be interpreted as a generalization of existing SRP beamformers.

\vspace*{0.5\baselineskip}%
\paragraph*{Criterion \numero1}
First, we consider the maximum source response criterion of~\eq\eqref{eq:criterion_no1}.
With $\m\Phi\xx(\omega) \shrt{=} \varPhi_{ss}(\omega) \, \m{h}\lt{s}(\omega) \, \mh{h}\lt{s}(\omega)$ of~\eq\eqref{eq:Phi_xx_rank1}, the source response in~\eq\eqref{eq:Phi_zz_sum} becomes
\begin{align}
	\srp^{(\m{x})}(\omega, \p)
	&=
	\varPhi_{ss}(\omega)\,
	\big\vert
	\mh{w}(\omega, \p) \, \m{h}\lt{s}(\omega)
	\big\vert^2 \,.
	\label{eq:Phi_zz_x}
\end{align}
By inserting the previous equation into~\eq\eqref{eq:criterion_no1}, dividing both sides of the inequality by $\Phi_{ss}(\omega)$ and taking the square root, we can reformulate the maximum source response criterion as
\begin{equation}
		\;
		\begin{gathered}
			\text{\textbf{Criterion \numero1}}
			\\[-3pt]
			\text{\textbf{(equivalent)}}
		\end{gathered}
		\quad\;
		\begin{gathered}
			\big\vert
			\mh{w}(\omega, \ps) \, \m{h}\lt{s}(\omega)
			\big\vert
			\geq
			\big\vert
			\mh{w}(\omega, \p) \, \m{h}\lt{s}(\omega)
			\big\vert
		\end{gathered}
		\,
		\label{eq:criterion_no1_beampattern}
\end{equation}
for all $\p$ and $\ps$.
In fact, this formulation shows that the maximum source response criterion is independent of the source signal.
Now we can insert the general beamformer formulation of~\eq\eqref{eq:w_general_fomulation} into the maximum source response criterion of~\eq\eqref{eq:criterion_no1_beampattern}:
\begin{align}
	\big|\alpha(\omega, \ps)\big|
	\big\vert
	\mh{h}\lt{s}\!(\omega) \, \m{A} \, \m{h}\lt{s}(\omega)
	\big\vert
	\geq
	\big|\alpha(\omega, \p)\big|
	\big\vert
	\mh{h}\!(\omega,\p) \, \m{A} \, \m{h}\lt{s}(\omega)
	\big\vert \,.
	\label{eq:criterion_no1_general_formulation}
\end{align}
In the \hyperref[apx:derivation_of_alpha]{Appendix}, it is shown that~\eq\eqref{eq:criterion_no1_general_formulation} -- and thus the maximum source response criterion -- is satisfied for every~$\p$ by choosing
\begin{align}
	\alpha(\omega, \p)
	&=
	\frac{ \zeta(\omega) }{ \sqrt{
			\mh{h}(\omega, \p) \, \m{A} \, \m{h}(\omega, \p) }} \,,
	\label{eq:alpha}
\end{align}
where $\zeta(\omega)$ is any positive, real-valued scalar.
With this, the general beamformer formulation of~\eq\eqref{eq:w_general_fomulation} results as
\begin{align}
	\m{w}(\omega,\p)
	&=
	\zeta(\omega) \,
	\frac{\m{A} \, \m{h}(\omega, \p)}{
		\sqrt{\mh{h}(\omega, \p) \, \m{A} \, \m{h}(\omega, \p) }
	} \,.
	\label{eq:w_GSRP_A}
\end{align}
This beamformer fulfills the maximum source response criterion of~\eq\eqref{eq:criterion_no1} for any Hermitian, positive-definite~$\m{A}$.

\vspace*{0.5\baselineskip}%
\paragraph*{Criterion \numero2}
In the next step, we must ensure that the constant noise response criterion of~\eq\eqref{eq:criterion_no2} is fulfilled.
To this end, we insert~\eq\eqref{eq:w_GSRP_A} into the noise response $\srp^{(\m{v})}(\omega, \p) = \mh{w}(\omega, \p) \, \m\Phi\vv(\omega) \, \m{w}(\omega, \p)$, and substitute it into~\eq\eqref{eq:criterion_no2}.
Thus,
\begin{align}
	\srp^{(\m{v})}(\omega, \p) 
	&=
	\zeta^2(\omega) \,
	\frac{
		\mh{h}(\omega, \p) \, \mh{A} \,\m\Phi\vv(\omega) \, \m{A} \, \m{h}(\omega, \p)
	}{
		\mh{h}(\omega, \p) \, \m{A} \, \m{h}(\omega, \p)
	} 
	= C
	\nonumber
	\\[-6pt]
	\label{eq:Phi_zz_v_A}
\end{align}
must hold for all $\p$.
It can be seen that this criterion is met by choosing $\m{A} \shrt{=} \mi\Phi\vv(\omega)$ as~\eq\eqref{eq:Phi_zz_v_A} then reduces to
\begin{align}
	\srp^{(\m{v})}(\omega, \p) 
	&=
	\zeta^2(\omega)
	\,.
	\label{eq:Phi_MVCNR_v}
\end{align}
The resulting beamformer thus has a \textit{constant noise response} over all $\p$.
With this, we have derived a GSRP beamformer which fulfills both beamformer criteria~\eq\eqref{eq:criterion_no1} and~\eq\eqref{eq:criterion_no2} for generic ATFs $\m{h}(\omega, \p)$ and generic invertible NCMs $\m\Phi\vv(\omega)$:
\begin{equation}
	\boxed{
		\;
		\m{w}\lt{MVCNR}(\omega, \p)
		=
		\zeta(\omega) \,
		\frac{ \mi\Phi\vv(\omega) \, \m{h}(\omega, \p) }{ 
			\sqrt{\mh{h}(\omega, \p) \, \mi\Phi\vv(\omega) \, \m{h}(\omega, \p)} } \,. \;
	}
	\label{eq:w_MVCNR}
\end{equation}
We call this beamformer \mrk{minimum variance constant noise response (MVCNR)} beamformer, where the naming is explained in the following section.

\subsection{Discussion of the MVCNR Beamformer}
\label{sec:MVCNR_interpretation}

\noindent
The MVCNR beamformer resembles the MVDR beamformer, however, with the main difference of the square root in the denominator in~\eq\eqref{eq:w_MVCNR}.
In this section, we will briefly identify the relation between MVCNR and MVDR.

When we steer the MVCNR beamformer towards the source position~$\ps$, its source response
according to~\eq\eqref{eq:Phi_zz_x} is
\begin{align}
	\srp\lt{\scriptsize MVCNR}^{(\m{x})}(\omega, \ps)
	&=
	\varPhi_{ss}(\omega) \;
	\zeta^2(\omega) \,
	\mh{h}\lt{s}(\omega) \, \mi\Phi\vv(\omega) \, \m{h}\lt{s}(\omega) \,.
	\label{eq:Phi_MVCNR_x}
\end{align}
From a comparison of~\eq\eqref{eq:Phi_zz_x} and~\eq\eqref{eq:Phi_MVCNR_x} we can infer that
\begin{align}
	\mh{w}\lt{MVCNR}(\omega, \ps) \, \m{h}\lt{s}(\omega)
	=	
	\zeta(\omega) \, \sqrt{ \mh{h}\lt{s}(\omega) \, \mi\Phi\vv(\omega) \, \m{h}\lt{s}(\omega) } \;.
	\label{eq:w_h_MVCNR}
\end{align}
This reveals a relevant difference from MVDR, which is based on the distortionless response constraint $\mh{w}(\omega, \ps) \, \m{h}\lt{s}(\omega) \shrt{=} 1$.
Hence, MVCNR is not distortionless but its output PSD is scaled to ensure that it reaches a global maximum at $\ps$ according to~\eq\eqref{eq:criterion_no1} and~\eqref{eq:criterion_no2}\footnote{%
Note that in special cases, other SRP beamformers might also satisfy the GSRP beamformer criteria of~\eq\eqref{eq:criterion_no1} and~\eqref{eq:criterion_no2}.
For instance, one can show that MVDR and also DS fulfills the criterion \numero1~\eqref{eq:criterion_no1} in the acoustic far field, and criterion \numero2~\eqref{eq:criterion_no2} if, in addition, the noise is spatially uncorrelated and homogeneous.
However, in general, this is not the case.}.
In turn, MVCNR is not scaling-invariant with regard to the noise power (in contrast to MVDR).
These aspects also become visible when rewriting~\eq\eqref{eq:w_MVCNR} as
\begin{flalign}
	&\m{w}\lt{MVCNR}(\omega, \p)
	\label{eq:w_MVCNR_MVDR}
	&
	\\[-2pt]
	&\;\shrt{=}
	\zeta(\omega) \sqrt{\mh{h}\!(\omega, \p) \mi\Phi\vv\!(\omega) \m{h}(\omega, \p)} \,
	\underbrace{\frac{ \mi\Phi\vv(\omega) \m{h}(\omega, \p) }{ 
		\mh{h}\!(\omega, \p) \mi\Phi\vv\!(\omega) \m{h}(\omega, \p) }}_{\m{w}\lt{MVDR}(\omega, \p)} \,.
	\nonumber
\end{flalign}
It is noteworthy that the MVCNR beamformer of~\eqref{eq:w_MVCNR} can also be derived via minimum variance optimization using Lagrange multipliers (similar to the derivation of the MVDR beamformer, e.g., in~\cite{VanTrees2002}) with \eq\eqref{eq:w_h_MVCNR} as linear constraint, i.e.,
\begin{gather}
	\min_{\m{w}(\omega, \p)} \, \mh{w}(\omega, \p) \, \m\Phi\vv(\omega) \, \m{w}(\omega, \p)
	\label{eq:MV_GSRP}
	\\[-4pt]
	\text{ subject to } \,
	\mh{w}(\omega, \p) \, \m{h}(\omega)
	=	
	\zeta(\omega) \, \sqrt{ \mh{h}(\omega) \, \mi\Phi\vv(\omega) \, \m{h}(\omega) } \,.
	\nonumber
\end{gather}
Due to this similarity to MVDR, however with a constant noise response instead of a distortionless signal response, we call it \textit{minimum variance constant noise response}.
Its output PSD according to~\eq\eqref{eq:Phi_zz} is
\begin{align}
	\srp(\omega, \p) &=
	\zeta^2(\omega) 
	\frac{ \mh{h}(\omega, \p)  \mi\Phi\vv(\omega) \, \m\Phi\yy(\omega) \, \mi\Phi\vv(\omega)  	\m{h}(\omega, \p) }{
		\mh{h}(\omega, \p) \, \mi\Phi\vv (\omega) \, \m{h}(\omega, \p) } .
	\nonumber
	\\[-6pt]
	\label{eq:Phi_MVCNR}%
\end{align}
Interestingly, a similar term was derived independently from this work via a deterministic maximum likelihood approach for binaural DOA estimation with hearing aids in~\cite{Zohourian2018}.

\subsection{Simplification for Uncorrelated and Homogeneous Noise}

\noindent
Under the assumption of spatially uncorrelated and homogeneous noise with 
$\m\Phi\vv(\omega) \shrt{=} {\sigma_v^2}(\omega) \, \m{I}$
according to~\eq\eqref{eq:Phi_vv_IID}, 
the MVCNR beamformer weights of~\eq\eqref{eq:w_MVCNR} simplify to
\begin{equation}
	\boxed{\;\;
		\m{w}\lt{NMF}(\omega, \p)
		=
		\frac{\zeta(\omega)}{{\sigma_v(\omega)}} \,
		\frac{ \m{h}(\omega, \p) }{ 
			\norm{ \m{h}(\omega, \p)} } \,.
		\;\;}
	\label{eq:w_NMF}
\end{equation}
We can recognize this as a unit-length \mrk{normalized matched filter (NMF)} \cite{Jan1995} that is scaled by the positive, real-valued factor $\zeta(\omega) / \sigma_v(\omega)$.
In \cite{Zohourian2018}, an un-scaled version of this beamformer is proposed as target beamformer amongst others in the context of binaural DOA estimation.
Its output PSD is
\begin{align}
	\srp\lt{NMF}(\omega, \p)
	&=
	\frac{\zeta^2(\omega)}{\sigma_v^2(\omega)} \,
	\frac{ \mh{h}(\omega,\p) \, \m\Phi\yy(\omega) \, \m{h}(\omega,\p) }{
		\mh{h}(\omega,\p) \, \m{h}(\omega,\p) }	\,.
	\label{eq:Phi_NMF}
\end{align}

\begin{figure}[!t]
	\centering
	\vspace*{-8pt}%
	\subfloat[SRP\;with\;MVCNR\;beamforming.]{%
		\includegraphics[scale=0.35, trim=5mm 0mm 20mm 8mm, clip]{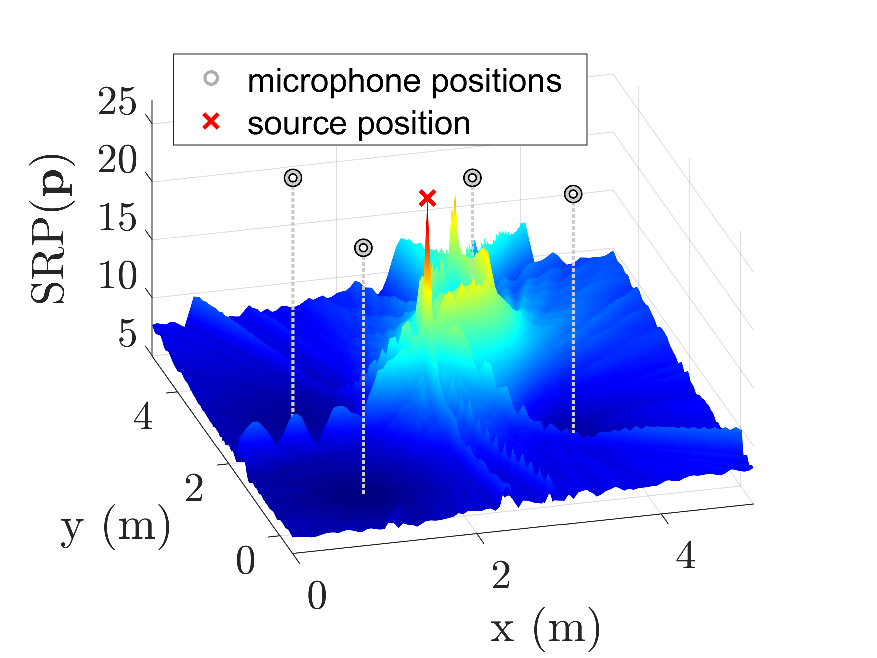}%
		\label{fig:SRP_MVCNR_nf}%
	}%
	\hfill
	\subfloat[SRP\;with\;MPCNR\;beamforming.]{%
		\includegraphics[scale=0.35, trim=2mm 0mm 20mm 8mm, clip]{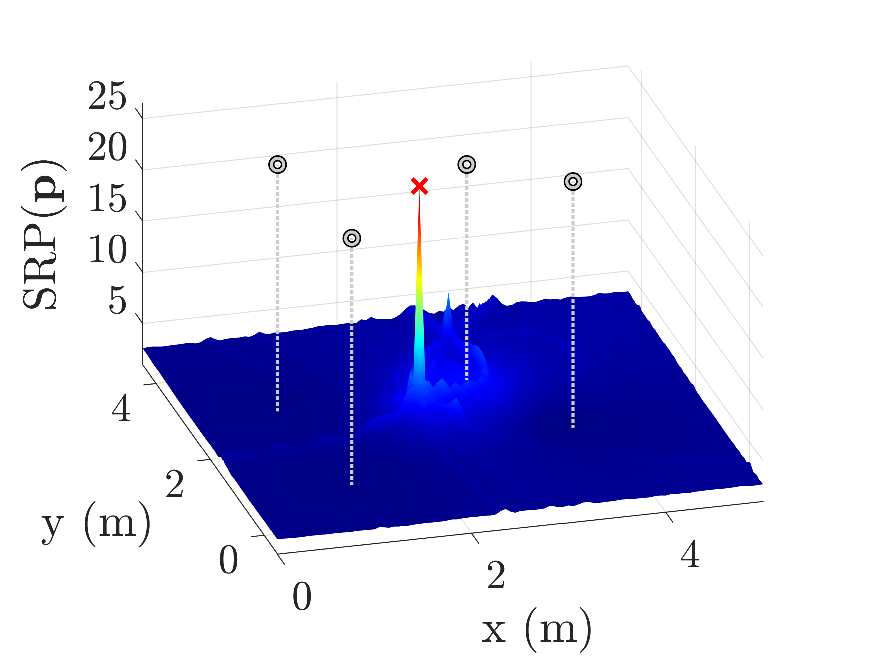}%
		\label{fig:SRP_MPCNR_nf}%
	}%
	\caption{%
		SRP maps when employing the acoustic free near-field model of~\eq\eqref{eq:free_near_field}
		with the proposed MVCNR~(a) or MPCNR~(b) beamformer.
		The simulated signals equal those used in \fig\ref{fig:DS_MVDR_nf}.
		In contrast to DS or MVDR, the maximum of the SRP maps with MVCNR or MPCNR coincide with the source position.
	}%
	\label{fig:SRP_MVCNR_MPCNR_nf}
\end{figure}

\subsection{GSRP Beamformer Derivation continued: The MPCNR}
\label{sec:further_GSRP_BFs}

\noindent
The MVCNR beamformer is not a unique solution for the GSRP beamformer design problem described in~\sect\ref{sec:GSRP_bf_design}.
In order to show this, we refer to the fact that MVDR and MPDR theoretically coincide $\m{w}\lt{MVDR}(\omega, \ps) \shrt{=} \m{w}\lt{MPDR}(\omega, \ps)$ without model errors, i.e., $\m{d}(\omega, \ps) \shrt{=} \m{h}(\omega, \ps)$, when they are steered towards~$\ps$~\cite{VanTrees2002, Ehrenberg2010}.
This property can be used to derive the minimum power pendant of the MVCNR beamformer.
Specifically, the respective beamformer weights can be determined by replacing the MVDR weights $\m{w}\lt{MVDR}(\omega, \p)$ in~\eq\eqref{eq:w_MVCNR_MVDR} by those of an MPDR beamformer, i.e.,
\begin{flalign}
	&\m{w}\lt{MPCNR}(\omega, \p)
	\label{eq:w_MPCNR}
	&
	\\[-2pt]
	&\;\shrt{=}
	\zeta(\omega) \sqrt{\mh{h}\!(\omega, \p) \mi\Phi\vv\!(\omega) \m{h}(\omega, \p)} \,
	\underbrace{\frac{ \mi\Phi\yy(\omega) \m{h}(\omega, \p) }{ 
			\mh{h}\!(\omega, \p) \mi\Phi\yy\!(\omega) \m{h}(\omega, \p) }}_{
		\m{w}\lt{MPDR}(\omega, \p)} \,.
	\nonumber
\end{flalign}
In accordance with MVCNR, we call this beamformer \textit{minimum power constant noise response} (MPCNR).
Similar as in~\eq\eqref{eq:MV_GSRP}, this solution can also be derived via minimum power optimization with $\min_{\m{w}(\omega, \p)} \, 
\mh{w}(\omega, \p) \, \m\Phi\yy(\omega) \, \m{w}(\omega, \p)$ subject to~\eq\eqref{eq:w_h_MVCNR}.
Moreover, one can show that the MPCNR beamformer also fulfills the GSRP beamformer condition of~\eq\eqref{eq:objective_Phi_zz}%
\footnote{The (not straightforward) proof is omitted here because this work focuses on the more practical MVCNR and NMF beamformers.}.
However, preliminary results showed that localization with MPCNR in practice is highly prone to model errors.
This also relates to the fact that MPDR is known to be less robust against versatile perturbations such as model mismatches compared to MVDR \cite{VanTrees2002, Ehrenberg2010}.
For this reason, we refrain from a systematic analysis of the MPCNR beamformer in the remainder of this paper to focus on the more robust variants MVCNR and NMF.

For comparison, \fig\ref{fig:SRP_MVCNR_MPCNR_nf} shows the SRP maps of the same simulated scenario as in \fig\ref{fig:DS_MVDR_nf} using MVCNR and MPCNR (with $\zeta(\omega)\shrt{=}1$) with the acoustic near-field model of~\eq\eqref{eq:free_near_field}.
Unlike with DS or MVDR, the SRP maps with MVCNR or MPCNR have a global maximum an the true source position.

\subsection{Frequency Weighting}
\label{sec:frequency_weighting}

\noindent
Whereas the previous sections dealt with the GSRP design of~\eq\eqref{eq:objective_Phi_zz} at each frequency $\omega$ independently, this section proposes and analyzes frequency weightings for the presented GSRP beamformers.
In the generic SRP framework (see \fig\ref{fig:SRP_structure}), the frequency weighting is introduced as positive, real-valued factor $\zeta^2(\omega)$ in~\eq\eqref{eq:SRP(p)}.
We can also recognize this scaling factor in the output PSDs of the GSRP beamformers in~\eq\eqref{eq:Phi_MVCNR} and~\eqref{eq:Phi_NMF}.
Hence, $\zeta^2(\omega)$ can be used to scale the contribution of each frequency to the broadband SRP result.
Below, we propose frequency weightings specifically for MVCNR which, however, apply in the same way for the simplified NMF.

With~\eq\eqref{eq:Phi_MVCNR_v} and~\eqref{eq:Phi_MVCNR_x}, the output PSD of the MVCNR beamformer in~\eq\eqref{eq:Phi_zz_sum}, when steered towards $\ps$, yields
\begin{align}
	\srp\lt{MVCNR}(\omega, \ps)
	&=
	\zeta^2(\omega) \,
	\big[
	\varPhi_{ss}(\omega) \;
	\mh{h}\lt{s}\!(\omega) \, \mi\Phi\vv(\omega) \, \m{h}\lt{s}(\omega) +
	1 \big] \,.
	\label{eq:Phi_MVCNR_max}
\end{align}
This can be rewritten with the trace operator $\tr(\cdot)$ and~\eq\eqref{eq:Phi_xx_rank1} as
\begin{align}
	\srp\lt{\scriptsize MVCNR}(\omega, \ps)
	&=
	\zeta^2(\omega) \,
	\big[
	\tr \big(
	\mi\Phi\vv(\omega) \, \m\Phi\xx(\omega) \big) +
	1 \big] \,,
	\label{eq:Phi_MVCNR_max_tr}
\end{align}
which is due to the circular shift invariance of the trace operator, i.e., 
$\mh{h}\lt{s}\!(\omega) \, \mi\Phi\vv(\omega) \, \m{h}\lt{s}(\omega) \shrt{=}
\tr(\mi\Phi\vv(\omega) \, \mh{h}\lt{s}\!(\omega) \, \m{h}\lt{s}(\omega))$.

\subsubsection{SNR weighting}
\label{sec:weighting_SNR}

Let us first consider a fixed, signal- and frequency-independent weight such as $\zeta\lt{SNR}^2(\omega) \shrt{=} {1}/{M}$, where the subscript refers to the signal-to-noise ratio (SNR).
Inserting into~\eq\eqref{eq:Phi_MVCNR_max_tr} yields
\begin{align}
	\srp\lt{\scriptsize MVCNR-SNR}(\omega, \ps)
	&=
	\frac{1}{M} \,
	\tr \big( \mi\Phi\vv(\omega) \, \m\Phi\xx(\omega) \big) \shrt{+} \frac{1}{M} \,.
	\label{eq:max_PSD_SNR}
\end{align}
This term can be interpreted as a lower-limited, narrowband-SNR-dependent weighting since the scaling depends on the noise-free SCM~$\m\Phi\xx(\omega)$ and the inverse NCM~$\mi\Phi\vv(\omega)$.
In particular, when assuming spatially uncorrelated and homogeneous noise according to~\eq\eqref{eq:Phi_vv_IID}, equation~\eq\eqref{eq:max_PSD_SNR} simplifies to
\begin{align}
	\!\!\srp\lt{\scriptsize MVCNR-SNR}(\omega, \ps)
	&\stackrel{\eqref{eq:Phi_vv_IID}}{=}
	\frac{\tr ( \m\Phi\xx(\omega) )}{ \tr ( \m\Phi\vv(\omega) ) } \shrt{+} \frac{1}{M}
	=
	\operatorname{SNR}(\omega) \shrt{+} \frac{1}{M} ,
	\label{eq:max_PSD_SNR_uncorr}
\end{align}
where $ \tr ( \m\Phi\vv(\omega) ) \shrt{=} M \, \sigma_{\!v}^2$ and the narrowband SNR is
\begin{align}
	\operatorname{SNR}(\omega)
	&=
	\frac{\Ex \{ \mh{x}(\omega) \, \m{x}(\omega) \}}{\Ex \{ \mh{v}(\omega) \, \m{v}(\omega) \}}
	=
	\frac{\tr ( \m\Phi\xx(\omega) )}{ \tr ( \m\Phi\vv(\omega) ) }
	\,.
	\label{eq:SNR}
\end{align}
The contribution of each frequency thus is directly related to its narrowband SNR.
This is also visualized in \fig\ref{fig:freq_weights} where $\srp\lt{MVCNR-SNR}(\omega, \ps)$ of~\eq\eqref{eq:max_PSD_SNR_uncorr} is plotted (dotted line) for spatially uncorrelated and homogeneous noise with various SNRs.

\vspace*{1\baselineskip}%
\subsubsection{Spectral flattening}
\label{sec:weighting_flat}

\begin{figure}[!t]
	\centering
	\includegraphics[width=0.80\linewidth, trim=0mm 0mm 0mm 0mm, clip]{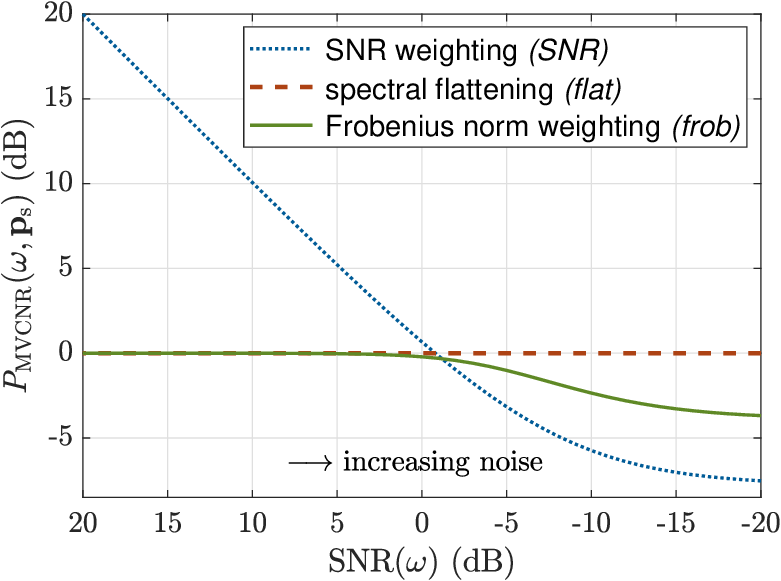}%
	\vspace*{-6pt}%
	\caption{
		Comparison of the GSRP beamformer output PSD $\srp\lt{MVCNR}(\omega,\ps)$ of~\eq\eqref{eq:Phi_MVCNR_max_tr} for all presented frequency weightings as a function of the narrowband SNR of~\eq\eqref{eq:SNR} with spatially uncorrelated and homogeneous noise.
		Simulation details: $\m\Phi\xx(\omega)$ is a randomly generated $M{\times}M$ Hermitian rank-one matrix with $M\shrt{=}6$; $\m\Phi\vv(\omega)$ as in~\eq\eqref{eq:Phi_vv_IID} and scaled according to $\operatorname{SNR}(\omega)$.%
	}%
	\label{fig:freq_weights}
\end{figure}

The signal or noise PSD levels can vary highly over frequencies, for instance, because of the frequency sparsity of speech or due to colored noise.
As a consequence, the SNR weighting might induce a highly non-uniform contribution of different frequencies where only a few single frequencies with highest narrowband SNR predominate the SRP result.
In order to prevent this, we propose a spectral flattening that equalizes the contribution of each frequency to the broadband SRP result.
This can be achieved, for instance, by enforcing
$\srp\lt{MVCNR}(\omega, \ps) \shrt{=} 1$ for all~$\omega$ regardless of the narrowband SNR (see~\fig\ref{fig:freq_weights}, dashed line).
With~\eq\eqref{eq:Phi_MVCNR_max_tr}, this directly yields the weighting
\begin{align}
	\zeta\lt{flat}^2(\omega)
	&=
	\big( \tr \big( \mi\Phi\vv(\omega) \, \m\Phi\xx(\omega) \big) \shrt{+} 1 \big)^{-1} \,.
	\label{eq:zeta_flat}
\end{align}

\subsubsection{Frobenius norm weighting}
\label{sec:weighting_frob}

We propose the following weighting as practical simplification of the spectral flattening:
\begin{align}
	\zeta\lt{frob}^2(\omega)
	&=
	\frac{\sigma_{\!v}^2(\omega)}{ {\norm{ \m\Phi\yy(\omega) }}_F } \,,
	\label{eq:frob_norm_weighting}
\end{align}
where 
${\sigma_{\!v}^2}(\omega) \shrt{=}
\Ex \{ \mh{v}(\omega) \, \m{v}(\omega) \} / M
\shrt{=}
\tr ( \m\Phi\vv(\omega) ) / M$
is the average noise power and
${\norm{ \m{A} }}_F \shrt{=} \sqrt{ \sum_{m=1}^{M} \sum_{n=1}^{N} |a_{mn}|^2}$ denotes the Frobenius norm of an $M \shrt{\times} N$ matrix $\m{A}$ with elements $a_{mn}$.
For spatially uncorrelated, homogeneous noise according to \eq\eqref{eq:Phi_vv_IID}, 
it can be shown that $\zeta\lt{frob}^2(\omega) \shrt{\approx} \zeta\lt{flat}^2(\omega)$ if $\operatorname{SNR}(\omega) \shrt{>} 0$\,dB, while $\zeta\lt{frob}^2(\omega)$ asymptotically approaches $\zeta\lt{flat}^2(\omega)/\sqrt{M}$ at noise-dominated frequencies.
In other words, each frequency has the same contribution at high SNR while highly noisy frequencies are attenuated up to a factor of $\sqrt{M}$.
This also becomes visible in \fig\ref{fig:freq_weights} (solid line).

Note that when applying the Frobenius norm weighting to the NMF beamformer, its output PSD~\eq\eqref{eq:Phi_NMF} simplifies to
\begin{align}
	\srp\lt{NMF-frob}(\omega, \p)
	&=
	\frac{ \mh{h}(\omega,\p)}{ \norm{\m{h}(\omega,\p)} } \, 
	\frac{ \m\Phi\yy(\omega) }{ {\norm{ \m\Phi\yy(\omega) }}_F } \, 
	\frac{ \m{h}(\omega,\p)}{ \norm{\m{h}(\omega,\p)} }	\,,
	\label{eq:NMF-frob}
\end{align}
which does not require a noise power estimate anymore.
One might observe a similarity between the Frobenius norm weight in the previous equation and the element-wise PHAT weighting~\cite{Knapp1976} used in SRP-PHAT.
In fact, the Frobenius norm weighting in~\eq\eqref{eq:NMF-frob} can be interpreted as a power-difference-preserving counterpart of the PHAT weighting, which yields an equalization of the SCM Frobenius norm over all frequencies while preserving the power differences of the microphone PSDs in the SCM.
As a consequence, the Frobenius norm weight coincides with the PHAT weighting (except for a scaling by $1/M$) if no power differences are observed between the microphones (e.g., in a compact microphone array in the free far-field of the source).

\begin{table*}[!t]
	\centering
	\caption{Overview of the proposed generalized SRP beamformers.}
	\label{tab:BF_overview}
	\vspace*{-1.5mm}%
	\begin{tabular*}{0.9\textwidth}{c@{\extracolsep{\fill}}c@{\extracolsep{\fill}}c}
		\textbf{SRP beamformer} & 
		\textbf{Beamformer weights} ${\displaystyle \m{w}(\omega,\p)}$ &
		\qquad \textbf{Output PSD} ${\displaystyle \srp(\omega,\p)}$
		\\[2pt] \hline
		\rule{0pt}{2.2\normalbaselineskip}%
		\begin{tabular}{@{}c@{}}
			Minimum variance\\constant noise response\\\mrk{(MVCNR)}
		\end{tabular}
		&
		\vspace*{-2pt}%
		${\displaystyle
			\zeta(\omega) \,
			\frac{ \mi\Phi\vv\!(\omega) \, \m{d}(\omega,\p) }{
				\sqrt{ \mh{d}\!(\omega,\p) \, \mi\Phi\vv\!\?(\omega) \, \m{d}(\omega,\p) } } }$
		\quad
		\begin{tabular}{@{}c@{}}
			cf.\\ \eq\eqref{eq:w_MVCNR}
		\end{tabular}
		&
		${\displaystyle
			\zeta^2(\omega) \,
			\frac{ \mh{d}\!(\omega,\p) \, \mi\Phi\vv\!\?(\omega) \, \m\Phi\yy(\omega) \, \mi\Phi\vv\!\?(\omega) \, \m{d}(\omega,\p) }{ 
				\mh{d}\!(\omega,\p)  \,\mi\Phi\vv\!\?(\omega)  \,\m{d}(\omega,\p) }}$
		\;
		\begin{tabular}{@{}c@{}}
			cf.\\ \eq\eqref{eq:Phi_MVCNR}
		\end{tabular}
		\\[14pt]
		\hline
		\rule{0pt}{2.1\normalbaselineskip}%
		\begin{tabular}{@{}c@{}}
			Normalized\\matched filter\\\mrk{(NMF)}
		\end{tabular}
		&
		${\displaystyle
			\frac{\zeta(\omega)}{ {\sigma_{v}(\omega)} } \,
			\frac{ \m{d}(\omega,\p) }{ \norm{ \m{d}(\omega,\p) }}}$%
		\quad
		\begin{tabular}{@{}c@{}}
			cf.\\ \eq\eqref{eq:w_NMF}
		\end{tabular}
		&
		${\displaystyle
			\frac{\zeta^2(\omega)}{ \sigma_{v}^2(\omega) } \,
			\frac{ \mh{d}\!(\omega,\p) \m\Phi\yy(\omega)  \m{d}(\omega,\p) }{ 
				\mh{d}(\omega,\p) \m{d}(\omega,\p) }}$
		\quad
		\begin{tabular}{@{}c@{}}
			cf.\\ \eq\eqref{eq:Phi_NMF}
		\end{tabular}
		\\[10pt] \hline
	\end{tabular*}
	\vspace*{-0.6\baselineskip}%
\end{table*}

\begin{table}[!t]
	\centering
	\caption{Proposed frequency weightings for the generalized SRP.}
	\label{tab:GSRP_freq_weighting}
	\vspace*{-1.5mm}%
	\begin{tabular*}{\linewidth}{c@{\extracolsep{\fill}}ccc}
		&
		\textbf{Method} &
		\textbf{Frequency weight} ${\displaystyle \zeta^2(\omega)}$
		&
		\\[1pt]
		\hline
		\rule{0pt}{1.7\normalbaselineskip}%
		&
		\begin{tabular}{@{}c@{}}%
			SNR weighting\\%
			\mrk{(SNR)}
		\end{tabular}
		&
		${\displaystyle	\frac{1}{M} }$
		&
		\\[10pt]
		&
		\begin{tabular}{@{}c@{}}%
			Spectral flattening\\%
			\mrk{(flat)}
		\end{tabular}
		&
		${\displaystyle
			\frac{1}{\tr ( \mi\Phi\vv(\omega) \, \m\Phi\yy(\omega) ) \shrt{-} M \shrt{+} 1 } }$%
		\quad\;\;
		\begin{tabular}{@{}c@{}}
			cf.\\ \eq\eqref{eq:zeta_flat}
		\end{tabular}
		&
		\\[12pt]
		&
		\begin{tabular}{@{}c@{}}%
			Frobenius norm weighting\\%
			\mrk{(frob)}
		\end{tabular}
		&
		${\displaystyle
			\frac{ {\sigma_{v}^2}(\omega) }{ {\norm{\m\Phi\yy(\omega)}}_F } }$
		\quad
		\begin{tabular}{@{}c@{}}
			cf.\\ \eq\eqref{eq:frob_norm_weighting}
		\end{tabular}
		\rule[-1.2\normalbaselineskip]{0pt}{0pt}
		&
		\\[9pt] \hline
	\end{tabular*}
	\vspace*{-0.4\baselineskip}
\end{table}

\subsection{GSRP Beamforming in Practice}
\label{sec:gsrp_in_practice}

\noindent
This section briefly discusses relevant aspects of the practical application of the above-presented GSRP method for acoustic source localization.

\vspace*{0.25\baselineskip}%
\subsubsection*{Acoustic models}

In practice, the ATFs~$\m{h}(\omega, \p)$ in the proposed beamformers are unknown and thus are replaced by an acoustic model $\m{d}(\omega, \p)$.
The respective beamformer weights and output PSDs are listed in \tab\ref{tab:BF_overview}.
Moreover, it is worth mentioning that the output PSDs of the GSRP beamformers are invariant with regard to a (frequency-dependent) complex-valued scaling of $\m{d}(\omega, \p)$.
As a consequence, the beamformer output PSD is identical regardless of whether applying ATFs or relative transfer functions (RTFs) as acoustic model $\m{d}(\omega, \p)$ since RTFs are scaled ATFs.

\vspace*{0.25\baselineskip}%
\subsubsection*{Frequency weightings}

In~\eq\eqref{eq:zeta_flat}, we introduced the spectral flattening which depends on the inverse NCM and the noise-free SCM.
While the NCM can be estimated directly, e.g., during pauses of the desired signal, the unknown noise-free SCM can not be estimated directly.
We propose to approximate
$\m\Phi\xx(\omega) \shrt{=} \m\Phi\yy(\omega) \shrt{-} \m\Phi\vv(\omega)$ using~\eq\eqref{eq:Phi_yy=Phi_xx+Phi_vv},
which allows us to rewrite
$ \tr ( \mi\Phi\vv(\omega) \, \m\Phi\xx(\omega) )$ as
$ \tr ( \mi\Phi\vv(\omega) \, \m\Phi\yy(\omega)  ) \shrt{-} M $.
The resulting frequency weight is listed in \tab\ref{tab:GSRP_freq_weighting}.

\subsubsection*{SNR-dependent NCM regularization}
\label{sec:regularization}

The proposed MVCNR beamformer (\tab\ref{tab:GSRP_freq_weighting})
indicates a noise-dependent linear transformation of the acoustic model vector~$\m{d}(\omega, \p)$ in the term $\mi\Phi\vv(\omega) \, \m{d}(\omega, \p)$.
In practice, this can be unfavorable in the low-noise case as $\mi\Phi\vv(\omega)$ might induce a considerable transformation of the model $\m{d}(\omega, \p)$ even though the noise component in the signals actually is negligible.
To this end, we propose a simple regularization of the NCM by means of adding a scaled identity matrix:
\begin{align}
	\m\Phi\lt{vv,reg}(\omega)
	&=
	\m\Phi\vv(\omega) + \epsilon\lt{reg} \, \sigma_{\!y}^2(\omega) \, \m{I} \,,
	\label{eq:Phi_vv_regularization}
\end{align}
where $\epsilon\lt{reg}$ is a small, positive, real-valued regularization factor and
$\sigma_y^2(\omega) \shrt{=} {\tr (\m\Phi\yy(\omega))}/{M}$ is the average microphone signal power.
For low noise with $\sigma_{\!v}^2(\omega) \shrt{\ll} \sigma_{\!y}^2(\omega)$, the regularized NCM approaches a scaled identity matrix (i.e., only a scaling and no transformation of $\m{d}(\omega, \p)$) whereas the NCM is virtually unmodified in the high-noise case.

\vspace*{0.25\baselineskip}%
\subsubsection*{Spatial aliasing, spatial sampling and signal bandwidth}

Spatial aliasing is a common problem of beamformer-based localization~\cite{Dmochowski2009}.
It becomes especially severe with distributed microphone setups as the aliasing frequency reduces with increasing microphone distance.
However, not only spatial aliasing needs to be considered but there is also a relation between the signal bandwidth and the spatial SRP resolution, i.e., the spatial resolution of the SRP grid search in~\eq\eqref{eq:grid_search}: A lower spatial SRP resolution reduces the upper frequency of the usable signal bandwidth~\cite{GarciaBarrios2021}.
Considering these effects might be crucial in practice, especially for setups involving larger microphone distances.
However, a detailed analysis of these aspects is beyond the scope of this paper.

\section{Evaluations}
\label{sec:evaluations}

\subsection{Simulation Setups}
\label{sec:simulation_setup}

\noindent
The proposed generalized SRP beamforming approaches are evaluated in comparison to the conventional SRP method in three scenarios with different microphone constellations with speech as source signal.

\subsubsection{Car setup -- Speaker position estimation with distributed microphones in a car}

The first scenario uses six distributed omnidirectional microphones in the car roof in a minivan (three seat rows with two distributed microphones per row) with reverberation time $T_{60} \shrt{\approx} 90$\,ms.
We evaluated five different speaker positions with frontal head orientation.
The microphones are elevated 30\,cm above the mouth of the speaker.
The geometry of the setup is plotted in \fig\hyperref[fig:setup_V]{\ref{fig:evaluation_setups}.1}.
For each speaker position, we generated six-seconds-long microphone signals (three seconds of stationary noise followed by three seconds of noisy speech) of two female and two male speakers with driving noise at different speeds.
The noise-free speech signals are simulated using clean speech snippets of the Clarity Speech Corpus~\cite{Graetzer2022}, which are convolved with measured in-car room impulse responses (RIRs) from~\cite{Mueller2024}. 
The RIRs were captured using a 
mouth simulator with frontal orientation.
The used dataset also contains multichannel driving noise recordings at stationary speeds between 0\,km/h and 150\,km/h~\cite{Mueller2024}.

\vspace*{0.25\baselineskip}%
\subsubsection{HA setup -- DOA estimation with binaural hearing aid microphone arrays in an office}

\begin{figure}[!t]
	\centering
	\vspace*{-2.5mm}%
	\subfloat[Car setup: Speaker position estimation\hspace*{4pt}\linebreak
	with six distributed microphones in a car\hspace*{4pt}\linebreak
	(microphones are elevated 30\,cm above the\hspace*{4pt}\linebreak
	source positions).]{%
		\includegraphics[height=0.435\linewidth, trim=0mm 0mm 0mm 0mm, clip]{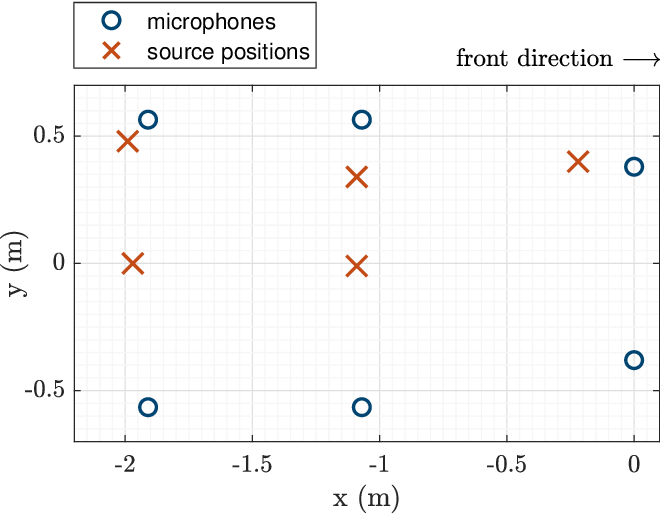}
		\label{fig:setup_V}
	}
	\hfill
	\subfloat[HA setup: DOA estimation with two behind-the-ear hearing aid devices mounted on a HATS in an office.]{%
		\includegraphics[height=0.435\linewidth, trim=0mm 0mm 0mm 0mm, clip]{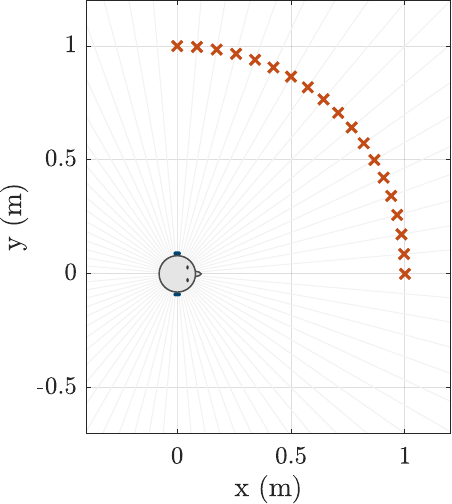}
		\label{fig:setup_UOL}
	}%
	\\
	\subfloat[UCA setup: DOA estimation with a uniform circular array (UCA) with 5\,cm diameter involving (a) five omnidirectional microphones or (b) five outward-oriented cardioid microphones in a simulated room.%
	]{%
		\includegraphics[height=0.4\linewidth, trim=-2mm -2mm -53mm 0mm, clip]{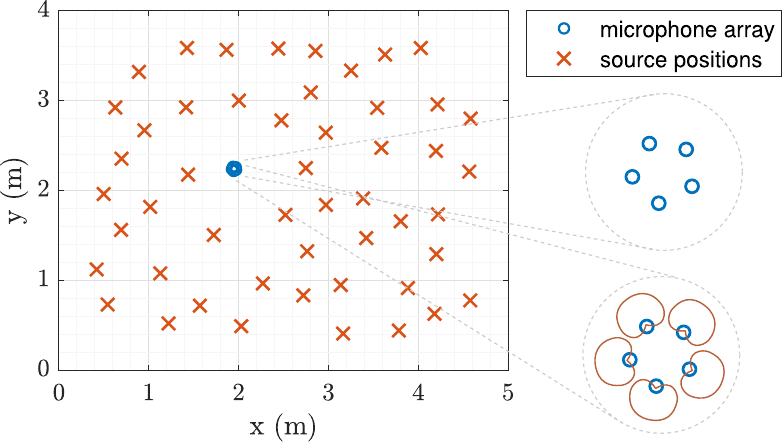}
		\label{fig:setup_DM}
	}
	
	\vspace*{-35mm}%
	\hspace*{58mm}%
	\scriptsize UCA(a) omnidirectional\\[-1pt]
	\hspace*{63mm} microphones\\
	
	\vspace*{10mm}
	\hspace*{50mm}%
	\scriptsize UCA(b) cardioid\\[-1pt]
	\hspace*{63.5mm} microphones
	\vspace*{14mm}%
	\caption{Geometry of the three evaluation setups.
	}%
	\label{fig:evaluation_setups}
\end{figure}

The second scenario uses two binaural behind-the-ear hearing aid (HA) devices mounted on a head and torso simulator (HATS) in an office room with reverberation time $T_{60} \shrt{\approx} 0.4$\,s.
Each HA device comprises three microphones with a distance of approximately 8\,mm (cf.~\cite{Kayser2009} for further details).
We evaluated 19 different speaker DOAs (elevation~0$^\circ$) between an azimuth of 90$^\circ$~(left) and 0$^\circ$~(front) in steps of 5$^\circ$ (see \fig\hyperref[fig:setup_UOL]{\ref{fig:evaluation_setups}.2}.).
For each speaker DOA, we generated six-seconds-long microphone signals (three seconds noise only plus three seconds noisy speech) of two female and two male speakers with multi-talker babble noise at different SNRs.
The noise-free speech signals are simulated using the same clean speech snippets as in the previous car setup, which are convolved with measured binaural room impulse responses (BRIRs) from~\cite{Kayser2009} (office~I).
Stationary, diffuse babble noise
was simulated according to \cite{Habets2008} with the NCM
$\m\Phi\lt{vv,\,diff}(k) \shrt{=}
\sum_{\theta \in [0 ... 360^\circ]}\,
\m{h}\lt{HRTF}(k, \theta) \, \mh{h}\lt{HRTF}(k, \theta) \,,$
where $k$ is the discrete Fourier transform frequency index and $\m{h}\lt{HRTF}(k, \theta)$ are the Fourier-transformed anechoic HRIRs of~\cite{Kayser2009} with DOA azimuth $\theta$ (in steps of 5$^\circ$).

\renewcommand*\thesubfigure{\arabic{subfigure}}
\renewcommand*\thesubfigure{\alph{subfigure}}

\vspace*{0.25\baselineskip}%
\subsubsection{UCA setup -- DOA estimation with a uniform circular array (UCA) in a simulated room}
\label{sec:uca_setup}

The third scenario uses a 5-microphone UCA with 5\,cm diameter involving (a)~omnidirectional microphones and (b)~outward-oriented cardioid microphones in a simulated room with reverberation time $T_{60} \shrt{=} 0.6$\,s.
This simulated setup was added in order to evaluate the applicability of the proposed GSRP methods in a (a)~typical setup of the conventional SRP (i.e., source in far field of the array without relevant source-position-dependent level differences), and in comparison to a (b)~setup involving an array of directional microphones (source in far field of the array, but relevant source-position-dependent level differences appear due to the microphone directivities).
We evaluated 50 randomly chosen speaker positions in the room (see \fig\hyperref[fig:setup_DM]{\ref{fig:evaluation_setups}.3}.).
For each speaker position, we generated six-seconds-long microphone signals (three seconds noise only plus three seconds noisy speech) of two female and two male speakers in an isotropic noise field (pink noise) at different SNRs.
The noise-free speech signals are simulated using the same clean speech snippets as in the car setup, which are convolved with simulated RIRs using~\cite{Habets2020, Allen1979}.
Stationary, diffuse pink noise 
was simulated with~\cite{Habets2008} where the spatial coherence function of the microphone array was computed according to~\cite{Elko2001} incorporating the different microphone directivities of setups (a) and (b).

\subsection{SRP Evaluation Method}

\noindent
We compared the performance of the proposed GSRP beamforming approaches against the conventional SRP by means of time-averaged SRP maps and the mean localization error.
In the \mrk{car setup}, the absolute localization error is the distance of the source position estimate~$\hat{\p}\lt{s}$ from the true source position~$\ps$, i.e., 
$E\lt{pos} \shrt{=} \Vert \, \hat{\p}\lt{s} \shrt{-} \ps \, \Vert$.
The absolute localization error in the \mrk{HA setup} and the \mrk{UCA setup} is the angular deviation
$E\lt{ang} \shrt{=} \vert \hat{\theta}\lt{s} \shrt{-} \theta\lt{s} \vert$
between the surce DOA estimate~$\hat{\theta}\lt{s}$ and the ground truth~$\theta\lt{s}$.
The mean localization error is computed by averaging the absolute localization error over time and over multiple speakers.

\subsubsection{Evaluated SRP algorithms}
As baseline, we considered the conventional \textit{SRP-PHAT} and the \textit{CSRP} without PHAT weighting (cf.\ \sect\ref{sec:SRP_framework}).
We compared the baseline against the proposed GSRP algorithms (cf.\ \tab\ref{tab:BF_overview}).
In particular, we considered \textit{MVCNR} in combination with the proposed frequency weightings \textit{SNR}, \textit{flat}, \textit{frob} (cf.\ \tab\ref{tab:GSRP_freq_weighting}), and the simpler \textit{NMF} (cf.\ \tab\ref{tab:BF_overview}) in combination with the \textit{SNR} and \textit{frob} weighting.
The \textit{flat} weighting was not considered in combination with the \mrk{NMF}.
This is because the practical benefit of \mrk{NMF} over \mrk{MVCNR} is that no NCM estimate is required which, however, is required by the \mrk{flat} weighting.

\vspace*{0.25\baselineskip}%
\subsubsection{Applied acoustic models}

As discussed in \sect\ref{sec:SRP_framework}, the baseline approaches \mrk{SRP-PHAT} and \mrk{CSRP} are based on the free far-field \mrk{(FF)} model of~\eq\eqref{eq:free_far_field}.
By contrast, GSRP can employ arbitrary acoustic models.
For the \textit{car setup} with distributed microphones, we considered the free near-field~\mrk{(NF)} model~$\m{d}\lt{nf}(k, \p)$ of~\eq\eqref{eq:free_near_field}.
In addition, we took into account the source directivity by exploiting knowledge of the frontal orientation of the speakers in the car.
The directivity of human speech can be modeled~\cite{Mueller2023}.
However, we used a dataset of measured, average speech directivities~\cite{Bellows2019, Leishman2021}.
The respective acoustic model for the GSRP methods in the car setup thus is determined as
$\m{d}\lt{GSRP}(k, \p) \shrt{=}
\m{d}\lt{nf}(k, \p) \odot \m{d}\lt{sd}(k, \p, 0^\circ)$,
where $\m{d}\lt{sd}(k, \p, 0^\circ)$ is the frequency-dependent average speech directivity of a speaker at position~$\p$ with orientation~0$^\circ$ (frontal orientation) towards all microphones and $\odot$ denotes the element-wise Hadamard product.
In the car setup, the acoustic models used for CSRP and GSRP thus have the same phase terms (cf.\ \eq\eqref{eq:free_far_field} and~\eqref{eq:free_near_field}) but differ in the level.

In the \textit{HA setup}, 
we used measured anechoic HRTFs of~\cite{Kayser2009} as acoustic model for the evaluated GSRP methods, i.e.,
$\m{d}\lt{GSRP}(k, \theta) \shrt{=} \m{h}\lt{HRTF}(k, \theta)$.
As discussed in \sect\ref{sec:SRP_extension}, the HRTFs comprise frequency-dependent interaural level and time differences that are caused by the acoustic head shadow.
Therefore, the HRTFs differ from the CSRP free-field model of~\eq\eqref{eq:free_far_field} in terms of both phase and level.
As an intermediate approach, we thus evaluated an additional, phase-corrected SRP-PHAT method.
It uses the phase of the HRTFs as acoustic model ${d}_{\text{HRTF-PHAT}, m}(k, \theta) \shrt{=} {h}_{\text{HRTF}, m}(k, \theta) / |{h}_{\text{HRTF}, m}(k, \theta)|$ instead of the free field model phase as proposed in~\cite{Rohdenburg2008}.

In the \textit{UCA(a) setup} involving omnidirectional microphone, 
we used the acoustic free far-field model $\m{d}\lt{ff}(\omega, \theta)$ of~\eq\eqref{eq:free_far_field} for both the conventional SRP methods and the GSRP methods.
In the \textit{UCA(b) setup} involving cardioid microphones, we considered the microphone directivities \mrk{(MD)} in addition to the far-field model for GSRP, i.e.,
$\m{d}\lt{GSRP}(\omega, \theta) \shrt{=} \m{d}\lt{ff}(\omega, \theta) \odot \m{d}\lt{MD}(\theta)$.
The directivity pattern of a cardioid microphone is given by 
$d_{\text{MD},m}(\theta) \shrt{=} 0.5 \, \big( 1 \shrt{+} \cos \angle (\phi_m, \theta) \big)$~\cite{Elko2001},
where $\phi_m$ reflects the orientation of microphone~$m$, and $\angle (\phi_m, \theta)$ denotes the intermediate angle between $\phi_m$ and $\theta$.

\vspace*{0.25\baselineskip}%
\subsubsection{SRP implementation details}
\label{sec:srp_implementation_details}

We implemented the SRP localization in the short-time Fourier transform (STFT) domain using Matlab.
All relevant implementation parameters can be found in~\tab\ref{tab:algo_details}.
The NCM was estimated by averaging over all instantaneous SCM estimates of the noise-only signal part (first three seconds) with
$\hat{\m\Phi}\vv(k) \shrt{=} 1/L\lt{n}
\sum_{l=1}^{L\lt{n}} \m{y}(k,l) \, \mh{y}(k,l)$
where $\m{y}(k,l)$ are the STFT signals of frequency bin $k$ and time frame~$l$, and $L\lt{n}$ is the number of frames of the noise-only signal part.
During the second, speech-plus-noise signal part, the SCM~$\hat{\m\Phi}\yy(k,l)$ was estimated in each frame by recursive smoothing over instantaneous SCM estimates
with smoothing time constant~$\tau\lt{sm} \shrt{=} $75\,ms (corresponding to smoothing factor $\alpha\lt{sm} \shrt{=} 0.2$).
The SRP map of~\eq\eqref{eq:SRP(p)} and the localization error was computed in each frame with speech activity.
Only frequencies above 100\,Hz were taken into account (by setting $\zeta^2(\omega) \shrt{=} 0$ in~\eq\eqref{eq:SRP(p)} for frequencies below 100\,Hz) since no relevant speech energy can be expected below.
Furthermore, frequencies in the car setup were upper limited to 4\,kHz because incorporating higher frequencies drastically increases the localization error of all evaluated SRP methods.
This is due to aliasing effects that mutually limit the usable signal bandwidth with a given spatial SRP map resolution as described in~\cite{GarciaBarrios2021}, especially if the source is close to the microphones.
As a consequence, we must limit the bandwidth to reduce undersampling of the SRP map.

\begin{table}[!t]
	\centering
	\caption{SRP implementation details of the evaluation setups.}
	\label{tab:algo_details}
	\vspace*{-1mm}%
	\begin{tabular*}{\linewidth}{ccc}
		\hline
		\rule{0pt}{1.2\normalbaselineskip}%
		Sampling frequency $f_s$   	& \multicolumn{2}{c}{ 16\,kHz }    \\
		\!\!\!\!STFT frame size\;$|$\;frame shift\;$|$\,window  \!\!\!\!	& \multicolumn{2}{c}{ 512\,samples\;$|$\;256\,samples\;$|$\;Hann}     \\
		SCM recursive smoothing $\tau\lt{sm}$\;$|$\;$\alpha\lt{sm}$  & \multicolumn{2}{c}{ 75\,ms $|$ 0.2 }       \\
		\!NCM regularization factor $\epsilon\lt{reg}$\! & \multicolumn{2}{c}{ 0.01 } \\[1pt]
		\hline
		\rule{0pt}{1.1\normalbaselineskip}%
		& \mrk{car setup} & \!\!\mrk{HA\shrt{/}UCA setup} \\
		SRP frequency boundaries   		& \!100\,Hz$\shrt{...}$4\,kHz  & \!\!100\,Hz$\shrt{...}$8\,kHz \\
		Spatial SRP resolution		& 5\,cm				& 5$^\circ$
		\\[1pt] \hline
	\end{tabular*}%
	\vspace*{-6pt}%
\end{table}

\subsection{Computational complexity}
\label{sec:computational_complexity}
\noindent
The computational complexity of the evaluated SRP methods differs in terms of the SRP beamforming method and the frequency weighting.
\tab\ref{tab:complexity} gives an overview of the most relevant computational differences between the beamforming methods 
\mrk{CSRP} (including \mrk{SRP-PHAT}), \mrk{NMF} and \mrk{MVCNR}.
While \mrk{NMF-SNR} requires a noise power estimate which scales the beamformer weights of~\eq\eqref{eq:w_NMF}, the \mrk{MVCNR} method requires the estimation and inversion of the full NCM.
These differences in computational complexity are particularly evident in real-time processing where the beamformer weights can be offline pre-computed for \mrk{CSRP} and \mrk{NMF} (except for a scalar multiplication for \mrk{NMF-SNR}).
By contrast, the \mrk{MVCNR} beamformer weights need to be updated in real time once a new NCM estimate is available.
\tab\ref{tab:proc_time} shows a comparison of the average processing time per STFT frame (the frame time period is 160\,$\mu$s) and the processing time factor relative to the \mrk{CSRP} method of our Matlab implementation of the \mrk{car setup}.

\begin{table}[!t]
	\centering%
	\caption{Comparison of the computational complexity of the\\presented SRP beamformers.}
	\label{tab:complexity}%
	\vspace*{-1mm}%
	\begin{tabular*}{0.97\linewidth}{c@{\extracolsep{\fill}}c@{\extracolsep{\fill}}c@{\extracolsep{\fill}}c}
		& \textbf{CSRP} & \textbf{NMF} & \textbf{MVCNR}
		\\[1pt]
		\hline
		\rule{0pt}{1.5\normalbaselineskip}%
		Noise estimation  & --    & 
		\begin{tabular}[c]{@{}c@{}}{\scriptsize\textit{NMF-SNR}}: noise power\\ {\scriptsize\textit{NMF-frob}}: --\end{tabular} & 
		full NCM
		\\[1pt]
		\hline
		\rule{0pt}{1.5\normalbaselineskip}%
		NCM inversion 	& --	& --	& yes
		\\[1pt]
		\hline
		\rule{0pt}{1.5\normalbaselineskip}%
		\begin{tabular}[c]{@{}c@{}}Noise-dependent\\ beamformer weights\end{tabular} & --    & \begin{tabular}[c]{@{}c@{}}{\scriptsize\textit{NMF-SNR}}: scalar only\\ {\scriptsize\textit{NMF-frob}}: --\end{tabular}         & yes
		\\[1pt]
		\hline
	\end{tabular*}
	\\
	\vspace*{2\baselineskip}
	\centering%
	\caption{Processing time of Matlab implementation (car setup).}
	\label{tab:proc_time}%
	\vspace*{-1mm}%
	\begin{tabular*}{0.97\linewidth}{c@{\extracolsep{\fill}}c@{\extracolsep{\fill}}c}
		\textbf{SRP method} & 
		\begin{tabular}[c]{@{}c@{}}\textbf{Processing time per frame}\\
			absolute / \% of frame time period
		\end{tabular} & 
		\textbf{Time factor}
		\\[1pt]
		\hline
		\rule{0pt}{1.1\normalbaselineskip}%
		CSRP       & 2.71\,µs \;/\; 1.70\,\%            & 1.00        \\
		SRP-PHAT   & 3.22\,µs \;/\; 2.01\,\%              & 1.19        \\
		NMF-SNR    & 4.47\,µs \;/\; 2.79\,\%             & 1.65        \\
		NFM-frob   & 4.57\,µs \;/\; 2.86\,\%             & 1.69        \\
		MVCNR-SNR  & 6.55\,µs \;/\; 4.09\,\%            & 2.42        \\
		MVCNR-flat & 6.80\,µs \;/\; 4.25\,\%            & 2.51        \\
		MVCNR-frob & 6.59\,µs \;/\; 4.12\,\%            & 2.43        \\ \hline
	\end{tabular*}
	\vspace*{-0.5\baselineskip}%
\end{table}

\subsection{Results}
\label{sec:results}

\subsubsection{Car setup}

\begin{figure}[!t]
	\centering
	\vspace*{-3mm}%
	\subfloat[SRP-PHAT \textit{(FF)}]{%
		\includegraphics[scale=0.19, trim=0mm 0mm 0mm 0mm, clip]{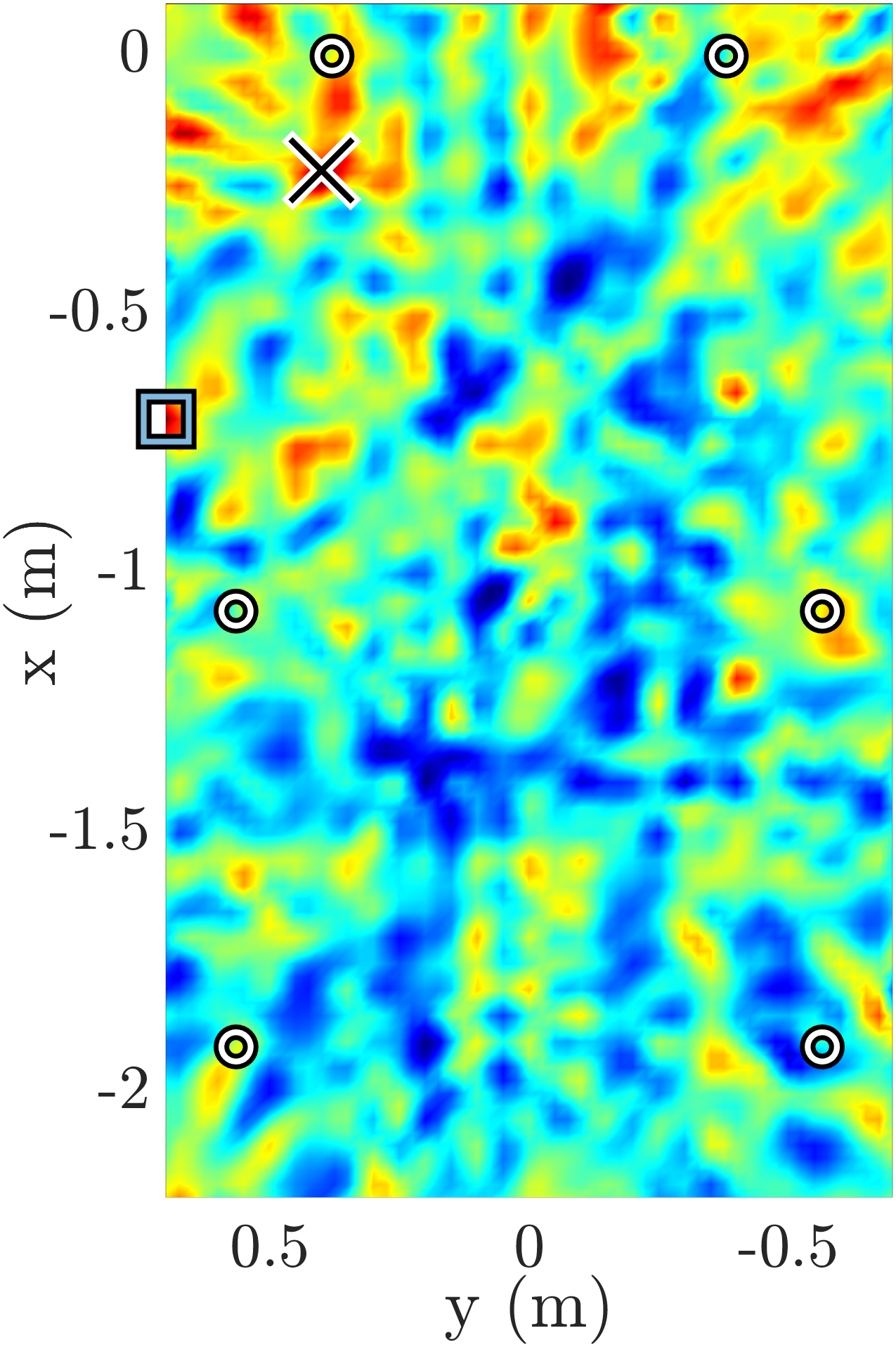}%
	}%
	\hfill
	\subfloat[\!MVCNR-frob\,\textit{(NF$\!$)\!}]{%
		\includegraphics[scale=0.19, trim=0mm 0mm 0mm 0mm, clip]{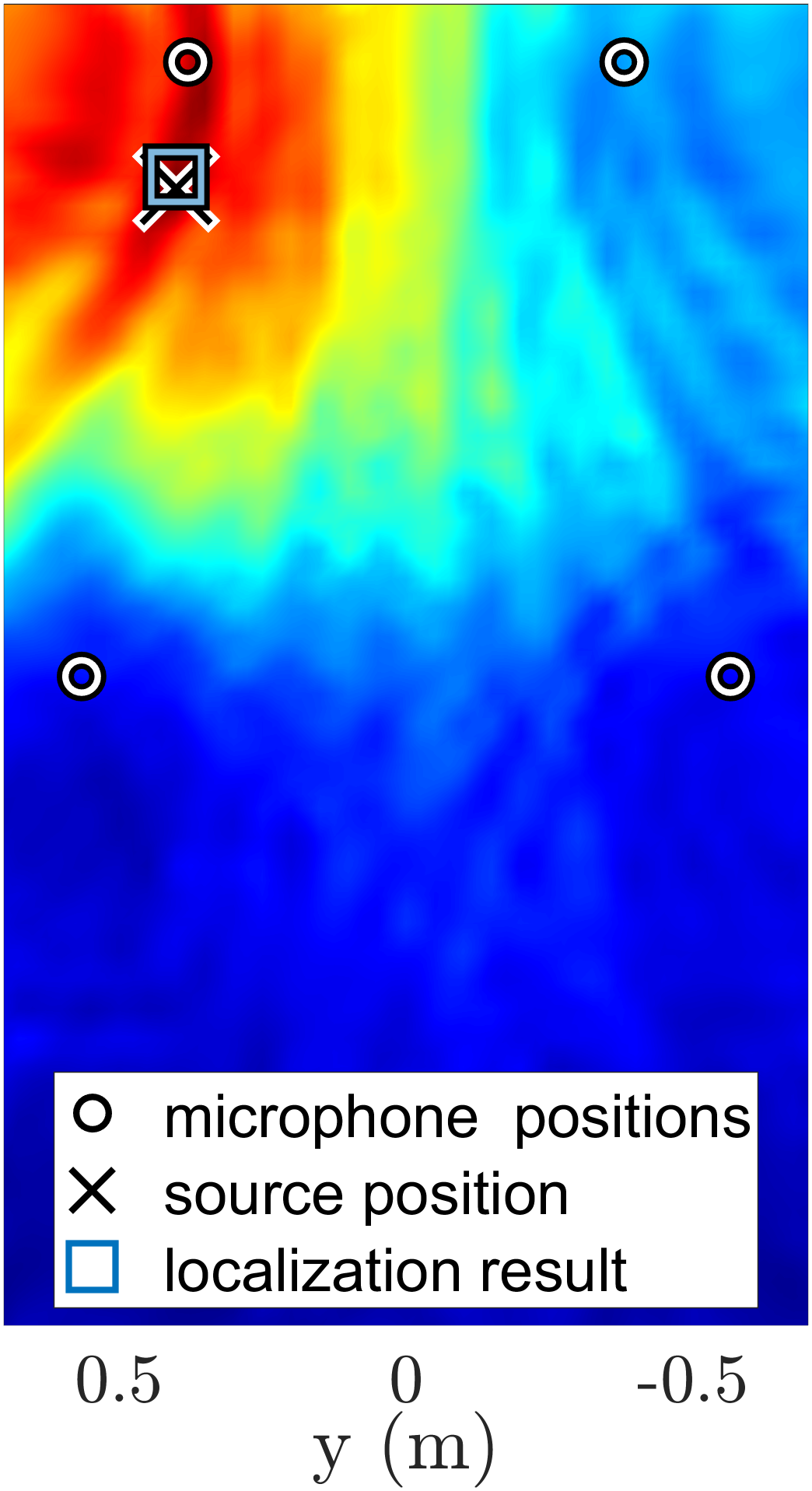}%
	}%
	\hfill
	\subfloat[NMF-frob \textit{(NF)}]{%
		\includegraphics[scale=0.19, trim=0mm 0mm 0mm 0mm, clip]{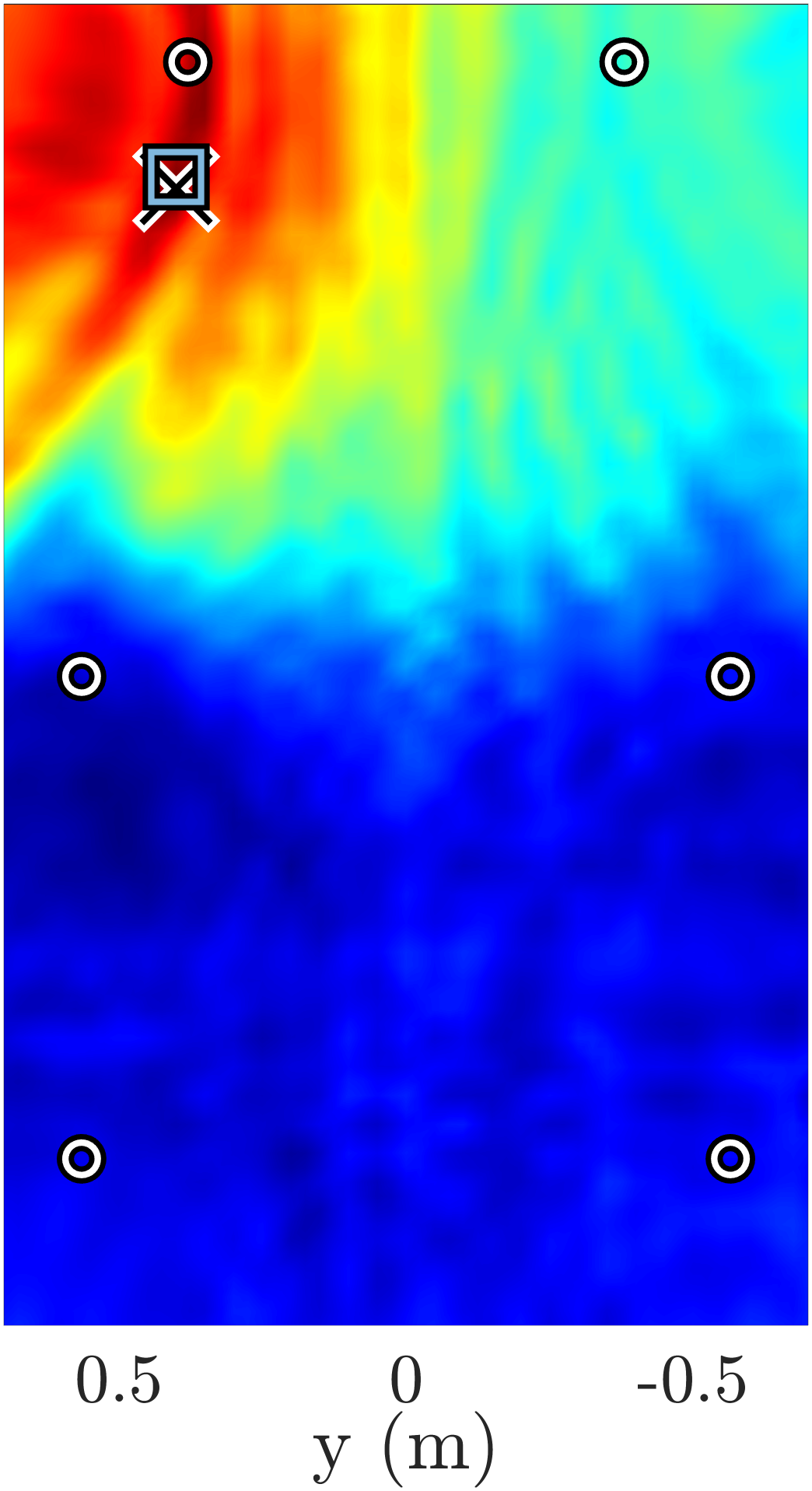}%
	}
	\subfloat{%
		\includegraphics[scale=0.19, trim=135mm 0mm 0mm 0mm, clip]{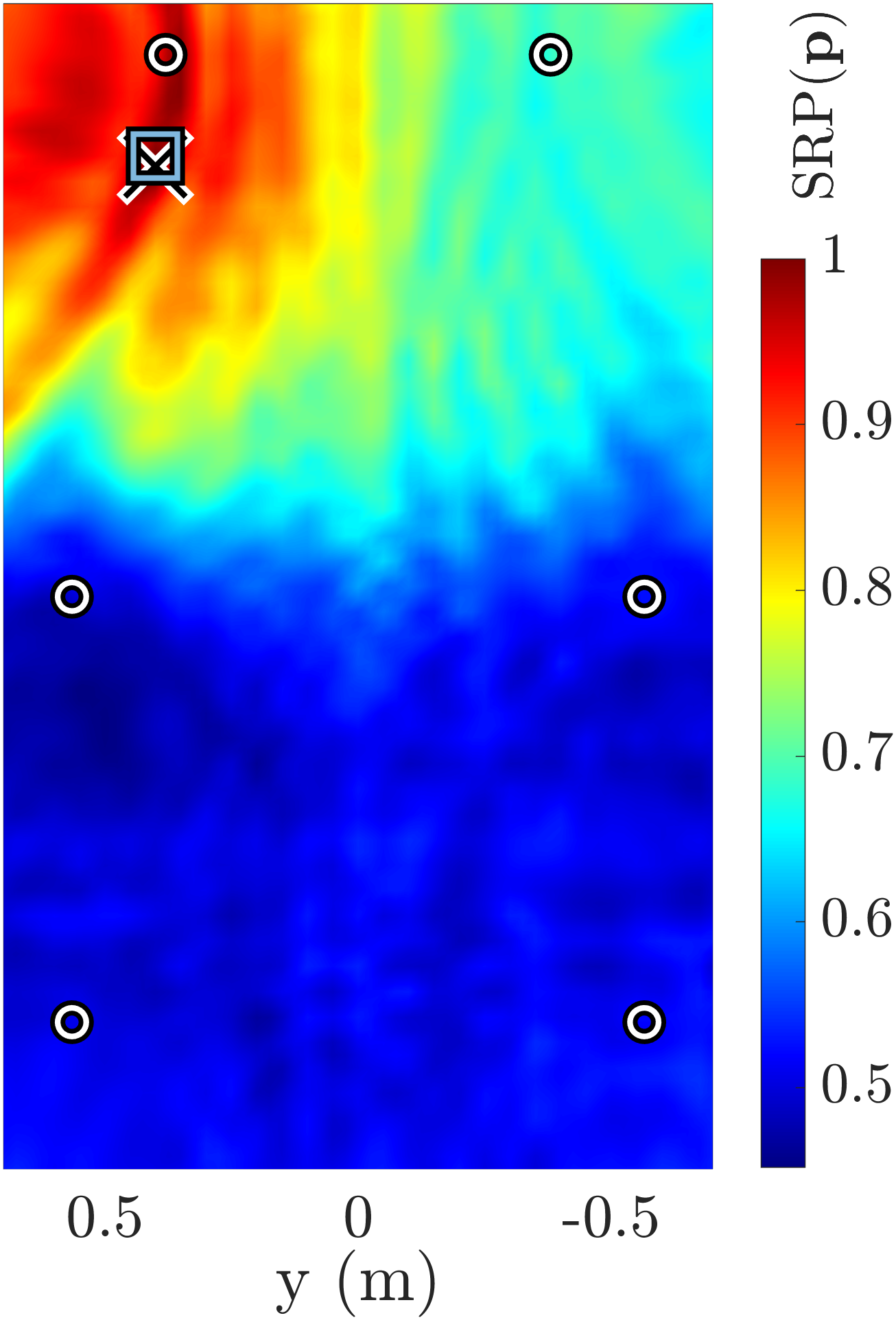}%
	}
	\caption{Car setup: Time-averaged SRP maps over 3\,s speech from the driver position at 120\,km/h driving speed (equivalent to $-$4\,dB SNR averaged over all microphones and 4\,dB SNR in the closest microphone, respectively).
		The acoustic model used in each evaluated SRP method is indicated in parentheses.
		(FF: free far-field model; NF: free near-field model.)%
	}
	\label{fig:SRP_map_V}
	\vspace*{6mm}%
	\includegraphics[scale=0.45, trim=0mm 0mm 0mm 0mm, clip]{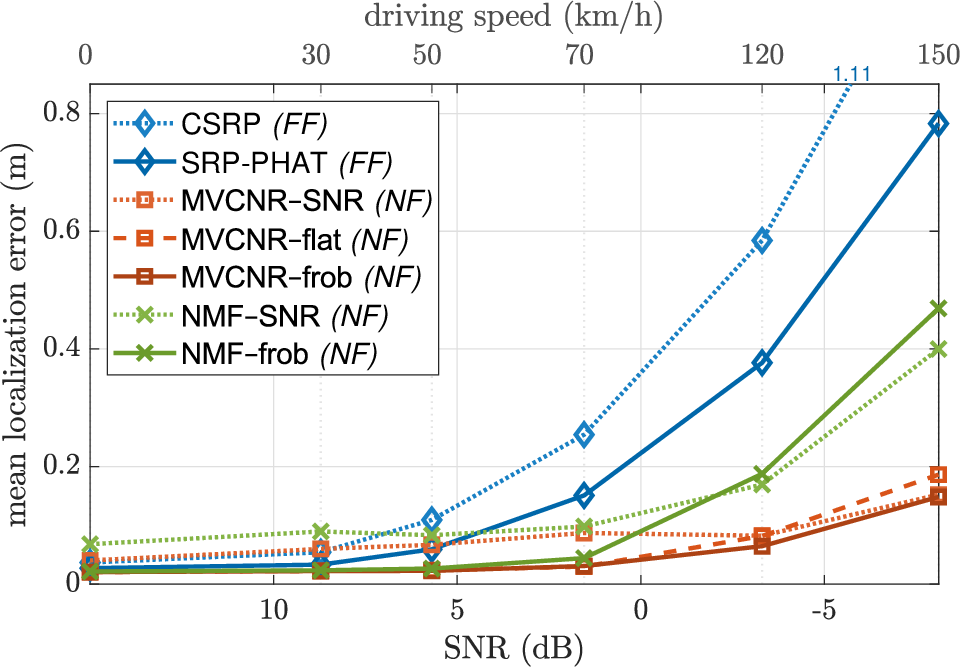}%
	\vspace*{-6pt}%
	\caption{Car setup: Mean localization error of all evaluated SRP methods over the microphone-averaged SNR.
		The respective driving speed is specified on the top axis.
		The acoustic model used in each evaluated SRP method is indicated in parentheses.
	}%
	\label{fig:mean_loc_error_car}
	\vspace*{7mm}%
	\includegraphics[scale=0.51, trim=0mm 0mm 3mm 0mm, clip]{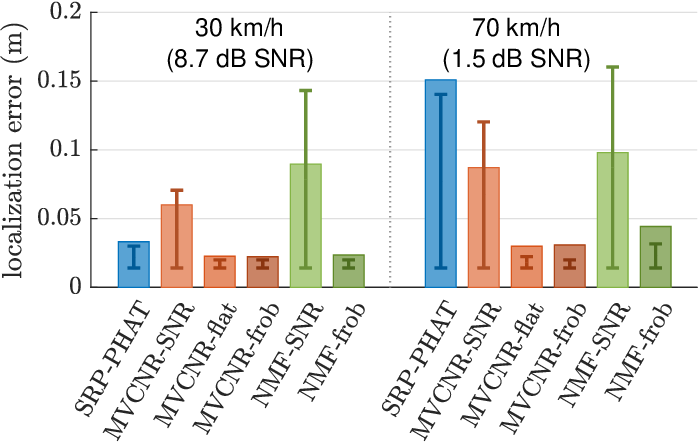}%
	\hfill
	\includegraphics[scale=0.51, trim=0mm 0mm 54mm 0mm, clip]{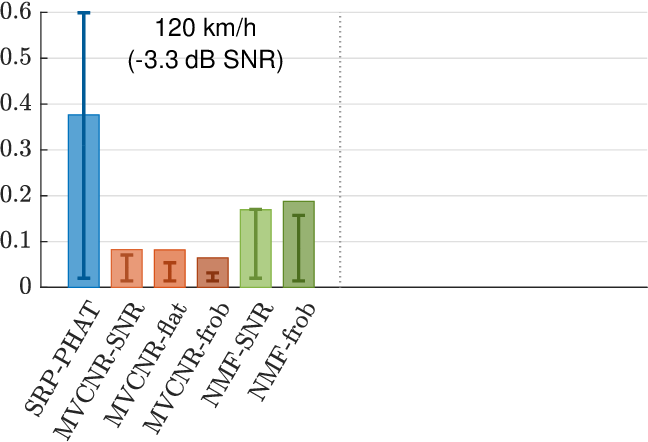}%
	\vspace*{-5pt}%
	\caption{Car setup: Mean localization error (wide bars) and upper and lower localization error quartiles at different driving speeds (different SNRs).
	}%
	\label{fig:mean_var_loc_error_car}
\end{figure}

\fig\ref{fig:SRP_map_V} shows time-averaged SRP heatmaps (i.e., the frame-wise SRP maps $\operatorname{SRP}(\p)$ of~\eq\eqref{eq:SRP(p)} are averaged over three seconds of noisy speech) of a speaker at the driver position at driving speed 120\,km/h for the conventional \mrk{SRP-PHAT} and the GSRP methods \mrk{MVCNR-frob} and \mrk{NMF-frob}.
The x- and y-axes indicate the Cartesian $x$ and $y$ coordinates of point~$\p$.
For better visualization, the SRP heatmaps are normalized to a value of one at the SRP maximum (square marker).
The \mrk{SRP-PHAT} map is very rough and has multiple sharp peaks.
Its maximum (square marker) does not coincide with the true source position.
By contrast, in the \mrk{MVCNR-frob} and \mrk{NMF-frob} maps, a larger area around the source position is elevated whereas areas in the SRP map which are distant from the source are consistently suppressed.
The \mrk{MVCNR-frob} map shows a better suppression of the region on the right-hand side of the source position (i.e., the front passenger seat in the car) compared to \mrk{NMF-frob}.
The maximum of both GSRP maps coincides with the source position whereas, in general, the peaks are less sharp than those in the \mrk{SRP-PHAT} map.

For a more systematic evaluation of all SRP variants, we computed the mean localization error (MLE).
To this end, the absolute localization error $E\lt{pos} \shrt{=} \Vert \, \hat{\p}\lt{s} \shrt{-} \ps \, \Vert$ was computed at each time frame with active speech and averaged over all frames, source positions and speakers.
In \fig\ref{fig:mean_loc_error_car}, the MLE (in meters) is plotted as a function of the driving speed and the respective average SNR over all microphones for all evaluated SRP methods.
At SNRs greater than 10\,dB, the MLE is low for all evaluated SRP methods.
At increasing speed (decreasing SNR), the performance between the different methods is increasingly diverging and the proposed GSRP approaches \mrk{MVCNR} and \mrk{NMF} in combination with any frequency weighting distinctly outperform the conventional methods \mrk{SRP-PHAT} and \mrk{CSRP}.
The MLE of \mrk{MVCNR} and \mrk{NMF} is comparable at positive SNRs whereas \mrk{MVCNR}, which includes a NCM estimate, outperforms \mrk{NMF} at negative SNRs and shows the best overall performance.
The \mrk{flat} and \mrk{frob} weighting with \mrk{MVCNR} perform similarly.
By contrast, the \mrk{SNR} weighting consistently performs worse at high SNRs whereas it is comparable or even preferable over \mrk{flat} and \mrk{frob} at negative SNRs.
At 150\,km/h, the MLE is reduced by a factor of five with \mrk{MVCNR-frob} compared to \mrk{SRP-PHAT}.

In addition to~\fig\ref{fig:mean_loc_error_car}, the upper and lower localization error quartiles are shown in comparison to the MLE for three different driving speeds in \fig\ref{fig:mean_var_loc_error_car} (note the different scaling of the y-axis for 120\,km/h).
This plot enables a better differentiability of the performance especially at low speed (high SNR) and, furthermore, it allows to assess the fluctuation of the localization error.
The \mrk{CSRP} method, which has the greatest overall MLE, is omitted for the purpose of better scaling.
At 30\,km/h, the MLE and the error quartiles of \mrk{SRP-PHAT} are comparable with those of \mrk{MVCNR} and \mrk{NMF} in combination with the \mrk{flat} and \mrk{frob} weighting.
However, at higher speeds, \mrk{SRP-PHAT} is clearly worse than the proposed GSRP methods.
In particular, the span of the error quartiles of \mrk{SRP-PHAT} is considerably higher which indicates a stronger fluctuation of the localization error.
A comparison of the proposed frequency weightings of the GSRP methods shows that the \mrk{SNR} weighting generates a greater MLE and error quartile span at high SNRs, whereas it is comparable at 120\,km/h with negative SNR.

\vspace*{0.25\baselineskip}
\subsubsection{HA setup}

\begin{figure}[!t]
	\centering
	\includegraphics[scale=0.425, trim=0mm 0mm 0mm 0mm, clip]{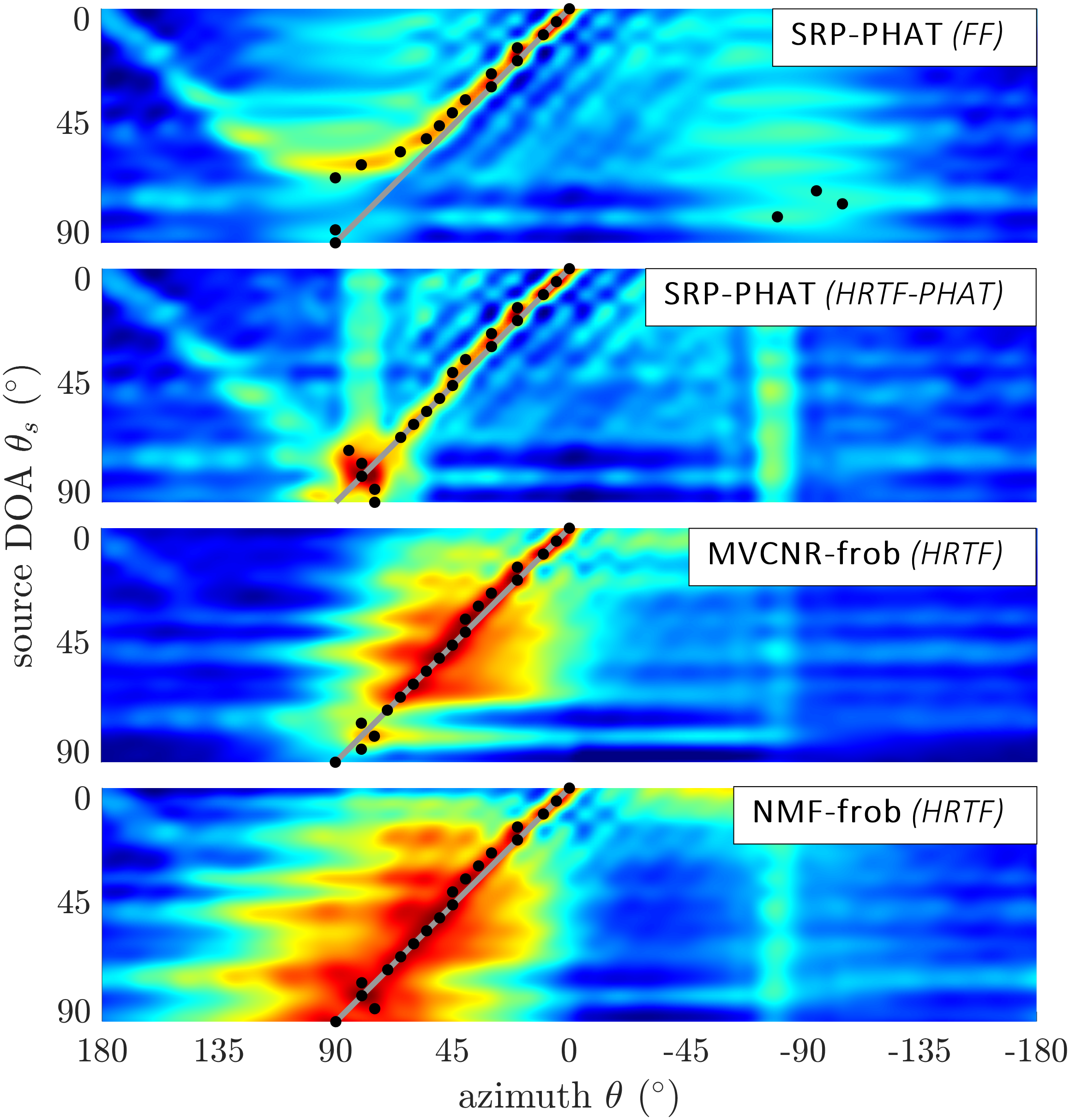}
	\vspace*{-3pt}%
	\caption{HA setup: Time-averaged, stacked DOA maps (one-dimensional DOA maps of several source directions $\theta\lt{s}$ are stacked on top of each other) of 3\,s speech snippets at 5\,dB SNR.
		Black dots denote the SRP maximum of each evaluated source DOA $\theta\lt{s}$.
		The acoustic model used in each SRP method is indicated in parentheses.
		(FF: free far-field model; HRTF: measured HRTFs; HRTF-PHAT: phase-transformed HRTF.)
	}
	\label{fig:SRP_map_UOL}
	\vspace*{5mm}
	\includegraphics[width=0.99\linewidth, trim=0mm 0mm 0mm 0mm, clip]{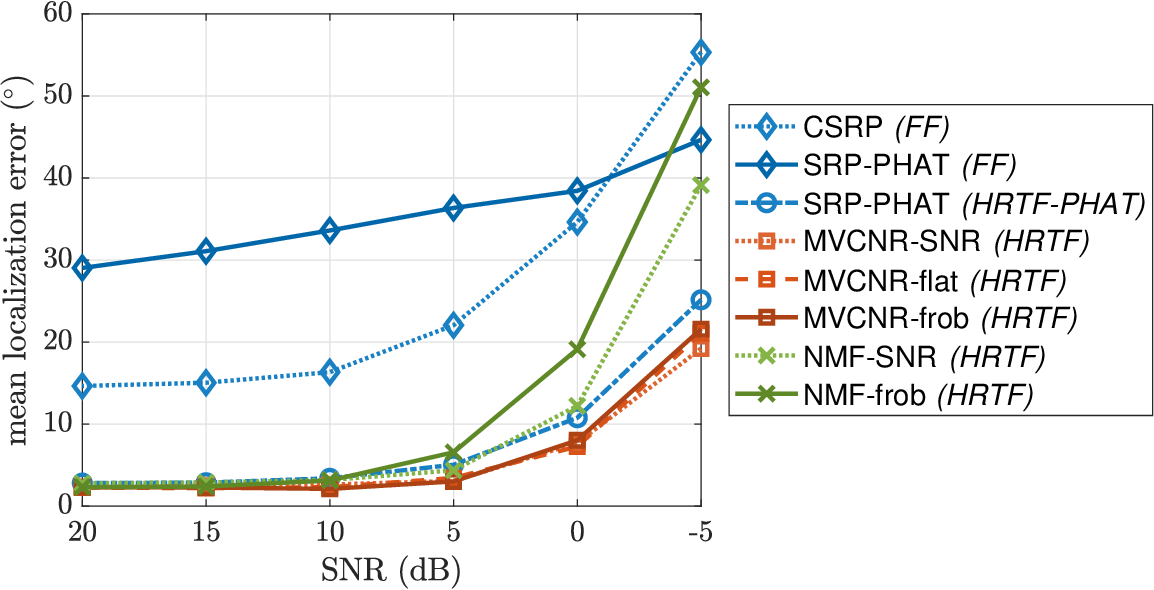}%
	\vspace*{-4pt}%
	\caption{HA setup: Mean localization error of all evaluated SRP methods over the SNR.
		The acoustic model used in each evaluated SRP method is indicated in parentheses.
	}
	\label{fig:mean_loc_error_HA}
\end{figure}

In the second evaluated setup, the source DOA is estimated.
This yields a one-dimensional SRP map over the azimuth $\theta$, which we call DOA map to distinguish from the two-dimensional SRP map of the previous evaluation setup.
We computed frame-wise DOA maps of 3\,s speech snippets at 5\,dB SNR for several source directions between $\theta\lt{s} \shrt{=} $90$^\circ$ and 0$^\circ$ with the conventional \mrk{SRP-PHAT} using a free far-field model \mrk{(FF)}, the modified \mrk{SRP-PHAT} method using the phase-transformed HRTF \mrk{HRTF-PHAT}, and the GSRP methods \mrk{MVCNR-frob} and \mrk{NMF-frob} using \mrk{HRTFs} as acoustic model.
In \fig\ref{fig:SRP_map_UOL}, the frame-wise DOA maps of each source direction were averaged over time and stacked on top of each other, which generates a two-dimensional DOA heatmap over several source directions~$\theta\lt{s}$.
The DOA map maximum of each source direction is marked by a black dot.
The gray line indicates the actual source DOA (ground truth).
While the \mrk{SRP-PHAT (FF)} map shows rather good results for frontal source DOAs between 45$^\circ$ and 0$^\circ$, the lateral DOAs are mislocalized.
By contrast, the \mrk{SRP-PHAT} using the \mrk{HRTF phase} is better able to resolve lateral source DOAs.
However, the localization of DOAs between 80$^\circ$ and 90$^\circ$ is still inaccurate.
The plotted GSRP maps show good localization performance over all evaluated DOAs as the SRP map maxima mostly coincide with the source DOAs (gray line).
However, the peaks of \mrk{MVCNR-frob} and especially \mrk{NMF-frob} are broader compared to \mrk{SRP-PHAT} whereas the GSRP methods, in particular \mrk{MVCNR-frob}, suppress the SRP map regions apart from the main peak to a greater extent.
This also reduces the front-back confusion compared to the \mrk{SRP-PHAT} maps.

\fig\ref{fig:mean_loc_error_HA} shows the MLE (in degree) of each evaluated SRP method, i.e., the average of the absolute angular deviation $E\lt{ang} \shrt{=} \vert \hat{\theta}\lt{s} \shrt{-} \theta\lt{s} \vert$ over all frames with active speech and all source DOAs.
The baseline \mrk{SRP-PHAT} and \mrk{CSRP} with the conventional free field model \mrk{(FF)} have a significantly higher MLE than all other methods already at high SNRs.
By contrast, the \mrk{SRP-PHAT} method involving the \mrk{HRTF phase} restores the localization capabilities and performs comparable to the proposed GSRP methods \mrk{MVCNR} and \mrk{NMF} at SNRs greater than 5\,dB.
At low SNR, the performance of \mrk{MVCNR} is clearly better compared to its simplified counterpart \mrk{NMF} which suffers from dominant noise.
The \mrk{SRP-PHAT} with \mrk{HRTF phase} is slightly worse than \mrk{MVCNR} at low SNRs.
In this setup, no significant differences can be seen between the proposed frequency weightings for \mrk{MVCNR}.
However, for the \mrk{NMF}, the \mrk{SNR} weighting is advantageous over the \mrk{frob} weighting at low SNRs.
The proposed \mrk{MVCNR} reduces the MLE by 25$^\circ$ to 30$^\circ$ compared to \mrk{SRP-PHAT (FF)} at all SNRs.%

\begin{figure}[!t]
	\vspace*{-4mm}
	\centering
	\subfloat[Omnidirectional microphones.]{%
		\begin{minipage}{0.49\linewidth}
			\includegraphics[scale=0.42, trim=15mm 16mm 20mm 5mm, clip]{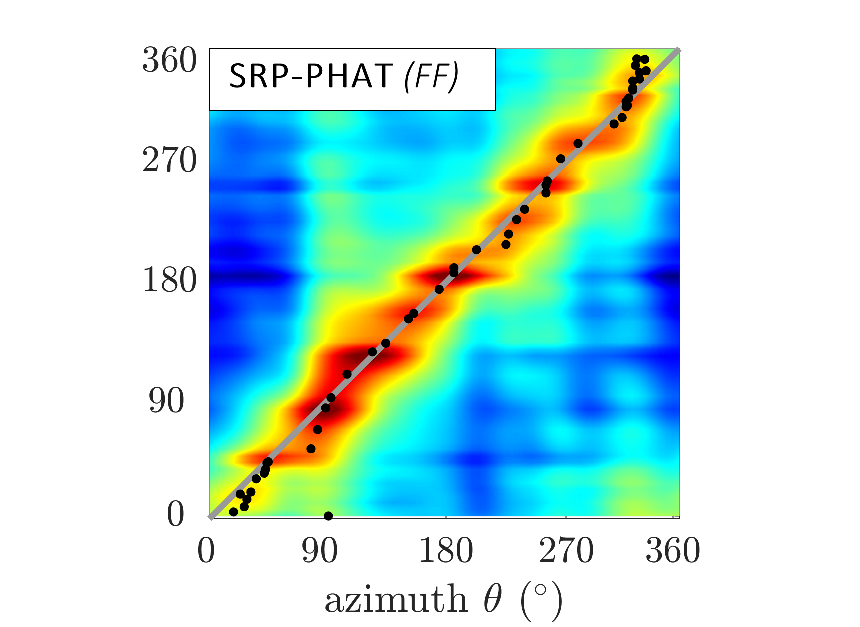}
			\\
			\includegraphics[scale=0.42, trim=15mm 16mm 20mm 5mm, clip]{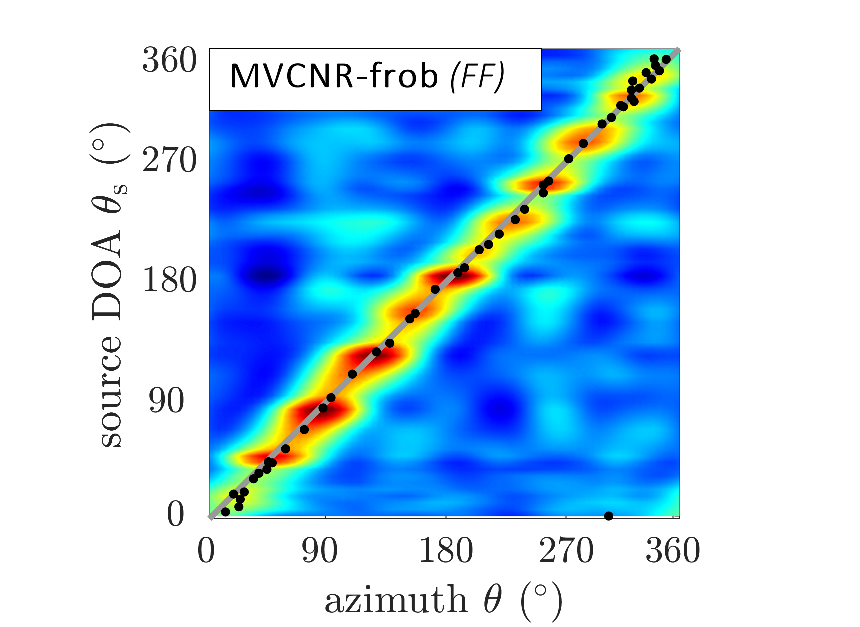}
			\\
			\includegraphics[scale=0.42, trim=15mm 0mm 20mm 5mm, clip]{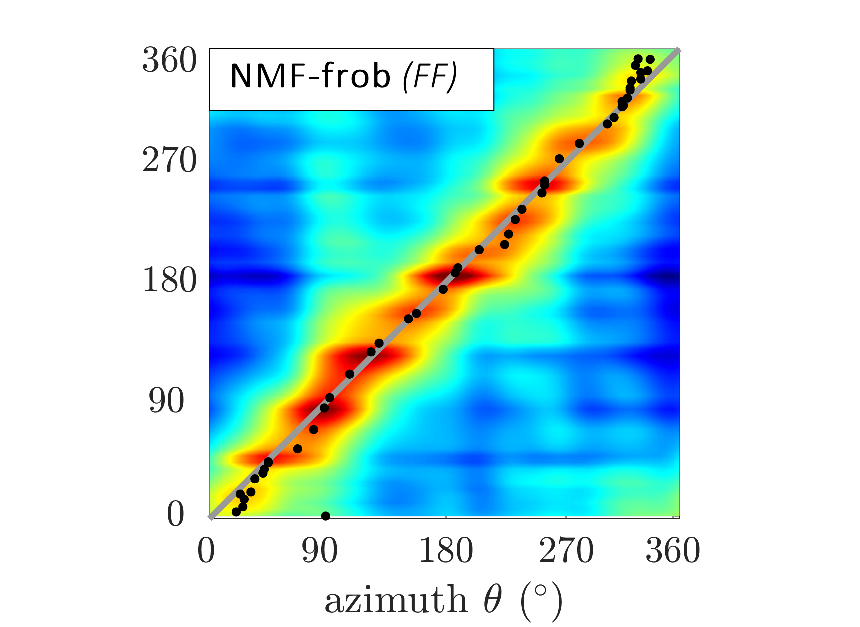}
		\end{minipage}
	}
	\subfloat[Cardioid microphones.]{%
		\begin{minipage}{0.49\linewidth}
			\includegraphics[scale=0.42, trim=15mm 16mm 20mm 5mm, clip]{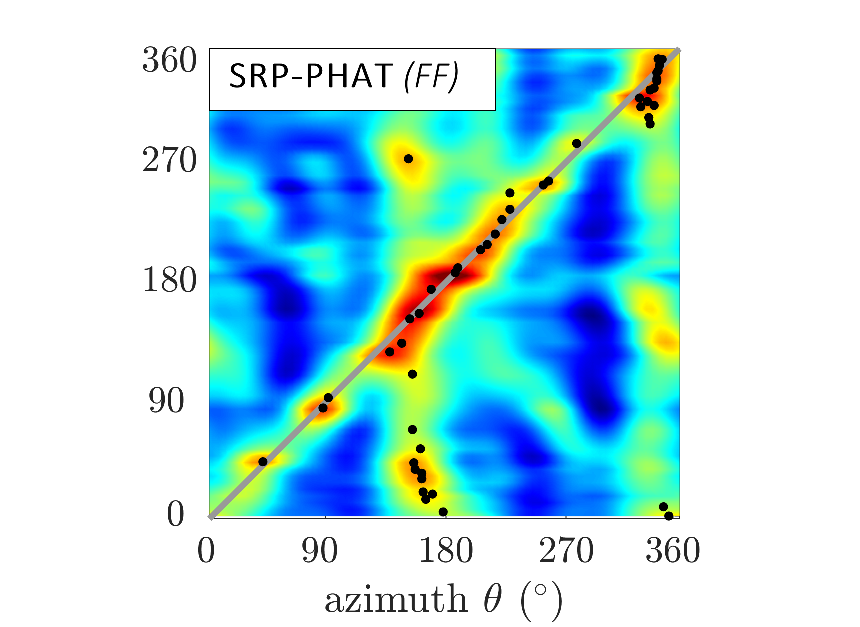}
			\\
			\includegraphics[scale=0.42, trim=15mm 16mm 20mm 5mm, clip]{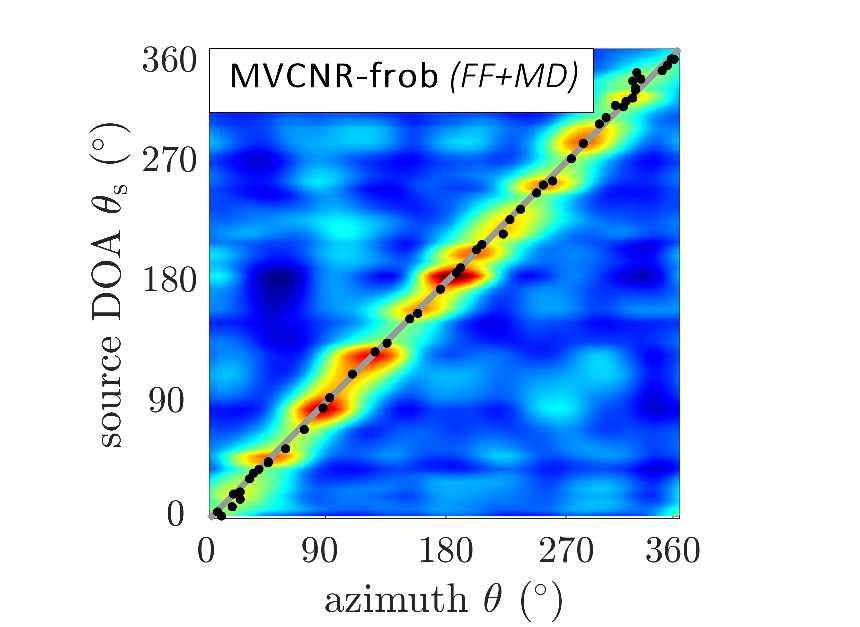}
			\\
			\includegraphics[scale=0.42, trim=15mm 0mm 20mm 5mm, clip]{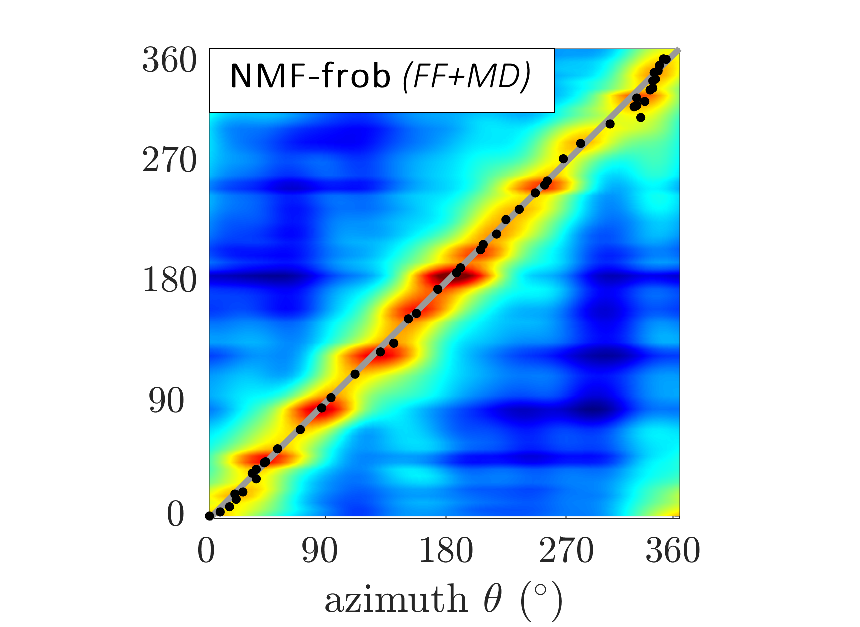}
		\end{minipage}
	}%
	\caption{UCA setup: Time-averaged, stacked DOA maps (one-dimensional DOA maps of several source directions $\theta\lt{s}$ are stacked on top of each other) of 3\,s speech snippets at 5\,dB SNR.
		Black dots denote the SRP maximum of each evaluated source DOA $\theta\lt{s}$.
		The acoustic model used in each SRP method is indicated in parentheses.
		(FF: free far-field model; FF+MD: free far-field model + cardioid microphone directivities.)
	}
	\label{fig:SRP_map_UCA}
\end{figure}

\vspace*{0.25\baselineskip}
\subsubsection{UCA setup}

In the third evaluation setup, we computed the time-averaged DOA maps of 3\,s speech snippets at 5\,dB SNR for all simulated source positions of the conventional \mrk{SRP-PHAT} and the GSRP methods \mrk{MVCNR-frob} and \mrk{NMF-frob}.
\fig\ref{fig:SRP_map_UCA} shows the stacked DOA maps of all evaluated source directions of the \mrk{UCA(a) setup} with omnidirectional microphones (left) in comparison to the \mrk{UCA(b) setup} with directional microphones (right).
The SRP maximum of each source DOA is marked by a black dot.
The gray line indicates the actual source DOA (ground truth).
In the \mrk{UCA(a) setup}, the SRP maps of all methods show comparably good results.
It is noticeable that the \mrk{SRP-PHAT} and \mrk{NMF-frob} map look almost identical and show slight DOA deviations at certain angles (e.g., close to $\theta\lt{s} \shrt{=} 0^\circ$ or $\theta\lt{s} \shrt{=} 90^\circ$).
The \mrk{MVCNR-frob} map shows good localization accuracy over all evaluated DOAs and regions apart from the source DOA are suppressed more consistently.
In the \mrk{UCA(a) setup} involving cardioid microphones, the performance of \mrk{SRP-PHAT} decreases significantly.
One can observe clear sidelobes in the DOA maps, e.g., around $\theta \shrt{=} 165^\circ$ or $\theta \shrt{=} 350^\circ$, which cause many DOA mislocalizations.
By contrast, the \mrk{MVCNR-frob} and \mrk{NMF-frob} DOA maps of the \mrk{UCA(b) setup} show similar or even slightly better accuracy compared to those of the \mrk{UCA(a) setup}.

\begin{figure}[!t]
	\vspace*{-2.5mm}%
	\centering
	\subfloat[Omnidirectional microphones.]{%
		\includegraphics[scale=0.46, trim=0mm 0mm 0mm 0mm, clip]{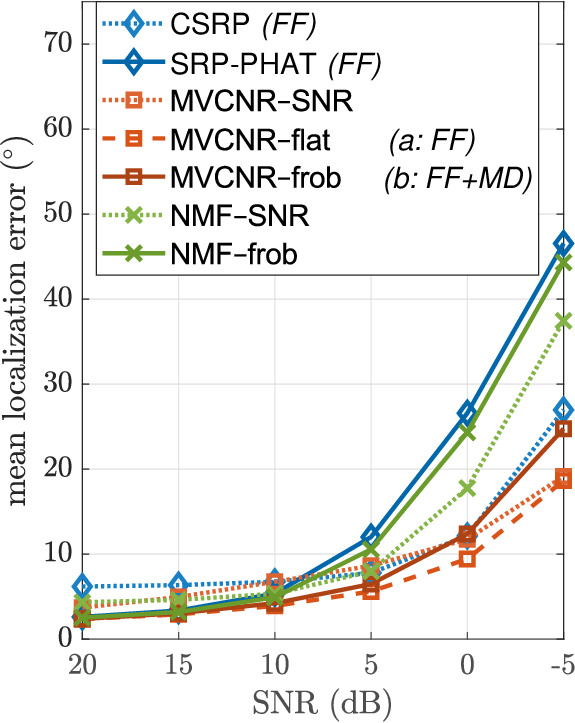}%
		\label{fig:mean_loc_error_UCAa}%
	}
	\hfil
	\subfloat[Cardioid microphones.]{%
		\includegraphics[scale=0.46, trim=0mm 0mm 0mm 0mm, clip]{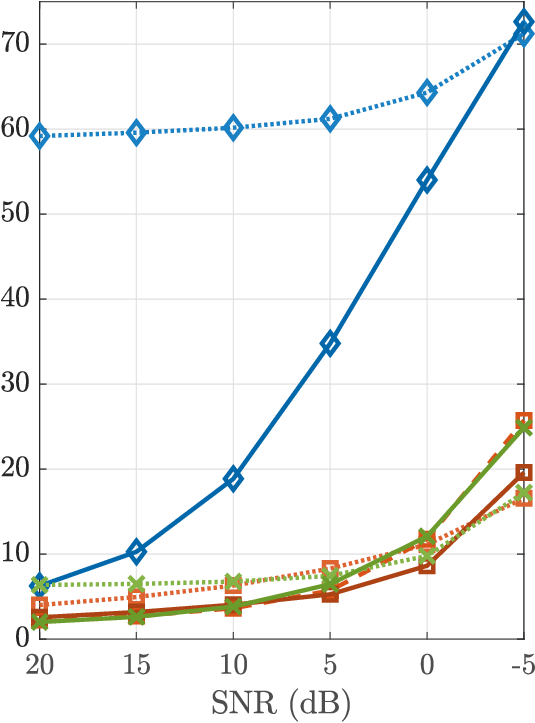}%
		\label{fig:mean_loc_error_UCAb}%
	}\\
	\vspace*{-54mm}%
	\!\hspace*{-29mm}%
	\,\tiny\rotatebox{90}{$\underbrace{\qquad\qquad\qquad\qquad}$}
	\vspace*{38mm}
	\caption{UCA setup: Mean localization error of all evaluated SRP methods over the SNR. 
		The acoustic model used in each evaluated SRP method is indicated in parentheses.
	}
	\label{fig:mean_loc_error_UCA}
\end{figure}

\fig\ref{fig:mean_loc_error_UCA} shows the MLE (in degree) averaged over all source DOAs of each evaluated SRP method for the \mrk{UCA(a) setup} (left) and \mrk{UCA(b) setup} (right) over various SNRs.
In the \mrk{UCA(a) setup} with omnidirectional microphones, all evaluated methods perform good at high SNR.
In particular, \mrk{SRP-PHAT}, \mrk{MVCNR} with \mrk{flat} and \mrk{frob} weighting, and \mrk{NMF-frob} show identical high accuracy for SNR greater than 10\,dB.
Whereas no relevant differences are observable between \mrk{SRP-PHAT} and \mrk{NMF-frob} at any SNR, the \mrk{MVCNR} methods clearly outperform \mrk{SRP-PHAT} and the \mrk{NMF} methods at SNRs lower than 5\,dB.
In the \mrk{UCA(b) setup} involving directional microphones, remarkable differences in the performance can be observed between the conventional methods \mrk{CSRP} and \mrk{SRP-PHAT} and the proposed GSRP methods.
The \mrk{CSRP} has extremely poor accuracy even at high SNR.
The MLE of \mrk{SRP-PHAT} at 20\,dB SNR is only slightly higher compared to the GSRP methods but increases significantly with lower SNR.
The performance of the GSRP methods is comparable, where \mrk{MVCNR-frob} shows a slightly lower MLE for positive SNRs and the methods involving the \mrk{SNR} weighting are preferable at negative SNR.
Compared to the \mrk{UCA(a) setup}, the performance of the \mrk{MVCNR} methods is similar whereas the \mrk{NMF} methods even perform better in the \mrk{UCA(b) setup} involving cardioid microphones.

\subsection{Discussion of Results}
\label{sec:discussion}

\noindent
The results show that the proposed generalized SRP methods \mrk{MVCNR} and \mrk{NMF} consistently outperform the conventional SRP baseline methods \mrk{SRP-PHAT} and \mrk{CSRP} in the presented evaluation setups.
In the \mrk{car setup}, the relevant difference between the conventional SRP methods and the proposed GSRP methods is that the latter exploits observed level differences in addition to TDOAs (acoustic free near-field model) whereas the conventional SRP methods only use TDOA information (free far-field model).
Especially at low SNRs, this exploitation of level information with GSRP significantly reduces the localization error in this setup.
In the \mrk{HA setup}, the proposed GSRP methods also use both microphone level and phase information for localization.
However, in this setup, the relevant improvement of GSRP over the conventional SRP, which uses the free far-field model, is the reduced TDOA mismatch between the acoustic model and the observed TDOAs by using measured HRTFs (or the HRTF phase) as acoustic model.
These HRTFs implicitly incorporate the shadowing effect of the head which causes significant deviations from the conventional free-field propagation, especially at lateral DOAs.
The \mrk{UCA setup} helps to further understand the benefit of the GSRP over the conventional SRP methods.
The \mrk{UCA(a) setup} with an array of omnidirectional microphones is a typical use case of the conventional SRP.
In this setup, the same acoustic model is applied to the GSRP methods as to the conventional SRP, which allows to asses the influence of the proposed GSRP frequency weightings and of incorporating the NCM estimate in the \mrk{MVCNR} method independently from the acoustic model.
As can be expected, the performance of \mrk{SRP-PHAT} and \mrk{NMF-frob} is almost identical in this scenario since no relevant microphone level differences are observable (cf. \sect\ref{sec:weighting_frob} for more details).
By contrast, the \mrk{MVCNR} beamformer is able to clearly improve the localization performance at low SNR.
The \mrk{UCA(b) setup} with an array of five directional microphones significantly impairs the localization performance of the conventional SRP method.
These results indicate that the phase information are less reliable in this setup with directional microphones.
A likely explanation for this is that phase differences of a microphone pair cannot be assessed if the source DOA is contrary to the orientation of one of the cardioid microphones.
For other source DOAs, phase information might be available at high SNR but they are prawn to phase distortions at lower SNR.
However, the results show that the GSRP method can compensate for these effects and restores or even slightly improves localization accuracy in the scenario involving directional microphones compared to the \mrk{UCA(a) setup}.
This is because considering the microphone directivity patterns in the acoustic model of the GSRP methods allows to additionally exploit the source-DOA-dependent level information of the directional microphones for localization.

In all simulated scenarios, \mrk{MVCNR} and \mrk{NMF} have similar performance at high SNR while \mrk{MVCNR} outperforms \mrk{NMF} at lower SNR.
This is intuitive since \mrk{NMF} is a simplification of \mrk{MVCNR} assuming spatially uncorrelated and homogeneous noise -- and this assumption is increasingly violated with increasing noise which, in fact, is not perfectly spatially uncorrelated and homogeneous.
Compared to \mrk{SRP-PHAT} and \mrk{CSRP}, the proposed GSRP methods not only have a lower MLE but also the fluctuation of the localization error is considerably lower especially at low SNR.
This is because the GSRP methods are less prone to TDOA estimation errors as they can additionally exploit level information which might indicate the (coarse) source position even with a noisy TDOA estimate.
Furthermore, the results compare the proposed GSRP frequency weightings \mrk{SNR}, \mrk{flat} and \mrk{frob}.
The \mrk{SNR} weighting scales each frequency depending on its respective narrowband SNR, which might cause a few single frequency bands with highest SNR to dominate the SRP result.
This property seems beneficial in the high-noise case with an SNR close to 0\,dB or lower, whereas the \mrk{flat} and the simpler \mrk{frob} weighting, which equalize the spectral contribution of each frequency, have a better performance at positive SNRs.

\section{Conclusion}
\label{sec:conclusion}

\noindent
In this paper, we have presented a generalization of the conventional SRP method that allows to exploit generic acoustic models and noise characteristics for acoustic sound source localization: the generalized steered response power (GSRP) method.
By using an appropriate acoustic model (including ATF or RTF measurements) for a given microphone setup, the proposed approach can jointly exploit the observed microphone level and phase differences to improve the localization performance compared to the conventional SRP which only assesses phase differences.
It has been shown that simply replacing the acoustic free far-field model of the conventional SRP method by other acoustic models is not optimal with known SRP beamformer methods, such as delay-and-sum or MVDR.
To this end, we propose a novel SRP beamforming design for localization using generic acoustic transfer functions and noise covariance matrices.
Based on this GSRP beamforming design, we have derived the \mrk{minimum variance constant noise response (MVCNR)} beamformer and its simplification for spatially uncorrelated and homogeneous noise -- the \mrk{normalized matched filter (NMF)}.
Furthermore, different frequency weightings for the presented beamformers have been proposed and analyzed.
These frequency weightings are suitable alternatives of the commonly used PHAT weighting for SRP as they, unlike PHAT, preserve the inter-microphone level differences that can be exploited by the GSRP method.
Realistic simulations of three different scenarios involving distributed microphones, hearing aid microphone arrays, and arrays with directional microphones under various noise conditions have shown that the mean localization error can be significantly reduced with the proposed methods compared to the conventional SRP.
In particular, MVCNR consistently performs well in all evaluated scenarios, whereas NMF suffers under highly noisy conditions.
The proposed SRP generalization is especially beneficial for setups where the conventional free far-field assumptions are violated to a greater extent and, moreover, the microphone level differences may contain relevant source location cues.
For instance, this is the case in setups with distributed microphones in the near field of the source, or setups involving acoustically shadowed or directional microphones.
However, also in a typical scenario of the conventional SRP with a speaker in the far field of a compact circular microphone array, the proposed GSRP methods were able to outperform the SRP-PHAT baseline.

{
\appendix
\label{apx:derivation_of_alpha}

\noindent
\textit{Derivation of $\alpha(\omega, \p)$ of~\eq\eqref{eq:alpha} based on the GSRP beamformer design criterion {\fontfamily{qpl}\textnumero}\,1~\eq\eqref{eq:criterion_no1_beampattern}:\;}
The matrix~$\m{A}$ in~\eq\eqref{eq:criterion_no1_general_formulation} is a Hermitian, positive-definite matrix and therefore can be decomposed with the Cholesky decomposition into $\m{A} \shrt{=} \m{L} \, \mh{L}$, where $\m{L}$ is a triangular matrix.
With this, \eq\eqref{eq:criterion_no1_general_formulation} becomes
\begin{align}
	&\big\vert\alpha(\omega, \ps) \,
	\mh{h}\lt{s}\!(\omega) \, \m{L} \, \mh{L} \, \m{h}\lt{s}(\omega)
	\big\vert
	\shrt{\geq}
	\big\vert\alpha(\omega, \p) \,
	\mh{h}\!(\omega,\p) \, \m{L} \, \mh{L} \, \m{h}\lt{s}(\omega)
	\big\vert .
	\label{eq:apx1}
\end{align}
Substituting 
$\m{h}'(\omega, \ps) \shrt{=} \mh{L} \, \m{h}\lt{s}(\omega)$ and 
$\m{h}'(\omega, \p) \shrt{=} \mh{L} \, \m{h}(\omega, \p)$ into~\eqref{eq:apx1} yields
\begin{align}
	\big\vert \alpha(\omega, \ps) \,
	\underbrace{
		{\m{h}'}\rmH\!(\omega, \ps) \, \m{h}'\!(\omega, \ps)
	}_{ {\Vert \m{h}'\!(\omega, \ps) \Vert}^2}
	\big\vert
	&\geq
	\big\vert \alpha(\omega, \p) \,
	{\m{h}'}\rmH\!(\omega, \p) \,
	\m{h}'\!(\omega, \ps) 
	\big\vert .
	\nonumber
	\\[-14pt]
	\label{eq:apx_signal_response_scalar_product}
\end{align}
Now, we search for an $\alpha(\omega, \p)$ and $\alpha(\omega, \ps)$, respectively, which ensure that this inequality
is true for all $\p$ and $\ps$.
To this end, we can use the Hermitian angle 
between two complex column vectors $\m{a}$ and $\m{b}$, i.e.,
$\cos\lt{H}(\m{a}, \m{b}) \shrt{=} |\mh{a} \, \m{b} | / (\norm{\m{a}} \, \norm{\m{b}} )$
with $\cos\lt{H}(\m{a}, \m{b}) \shrt{\in} [0, 1]$,
to rewrite \eq\eqref{eq:apx_signal_response_scalar_product} as
\begin{align}
	&\big\vert \alpha(\omega, \ps) \big\vert\,
	{\Vert \m{h}'\!(\omega, \ps) \Vert}^2
	\label{eq:apx_signal_response_cosH}
	\\
	&\;\,\geq
	\big\vert\alpha(\omega, \p)\big\vert \,
	\cos\lt{H} \big( 
	\m{h}'\!(\omega, \p), \,\m{h}'\!(\omega, \ps) \big) \,
	{\Vert \m{h}'\!(\omega, \p) \Vert} \,
	{\Vert \m{h}'\!(\omega, \ps) \Vert} .
	\nonumber
\end{align}
When dividing both sides of~\eq\eqref{eq:apx_signal_response_cosH} by ${\Vert \m{h}'\!(\omega, \ps) \Vert}$,
we can see that the inequality holds for 
\begin{align}
	\alpha(\omega, \p)
	&=
	\frac{\zeta(\omega)}{ 
		\Vert \m{h}'(\omega, \p) \Vert
	} \,, \,
	\text{ and }
	\alpha(\omega, \ps)
	=
	\frac{\zeta(\omega)}{ 
		\Vert \m{h}'(\omega, \ps) \Vert
	} \,,
	\label{eq:apx_alpha}
\end{align}
respectively, because \eq\eqref{eq:apx_signal_response_cosH} reduces with~\eq\eqref{eq:apx_alpha} to
\begin{align}
	1
	&\geq
	\cos\lt{H} \big( 
	\m{h}'\!(\omega, \p), \,\m{h}'\!(\omega, \ps) \big) \,,
\end{align}
which is true for all $\p$.
In~\eq\eqref{eq:apx_alpha}, $\zeta(\omega)$ is a positive, real-valued scalar.
Finally, when substituting $\m{h}'(\omega, \p) \shrt{=} \mh{L} \, \m{h}(\omega, \p)$, the found solution for $\alpha(\omega, \p)$ in~\eq\eqref{eq:apx_alpha} becomes
\begin{align}
	\alpha(\omega, \p)
	&=
	\frac{ \zeta(\omega) }{ 
		\big\Vert \mh{L} \, \m{h}(\omega, \p) \big\Vert
	}
	=
	\frac{ \zeta(\omega) }{ \sqrt{
			\mh{h}(\omega, \p) \, \m{L} \,
			\mh{L} \, \m{h}(\omega, \p) }} \,,
\end{align}
which can be rewritten by re-composing $\m{L} \, \mh{L} \shrt{=} \m{A}$ as
\begin{align}
	\alpha(\omega, \p)
	&=
	\frac{ \zeta(\omega) }{ \sqrt{
			\mh{h}(\omega, \p) \, \m{A} \, \m{h}(\omega, \p) }} \;.
	\label{eq:apx_alpha_final}
\end{align}

}


\pagebreak

\end{document}